\documentclass[10pt]{article}

\pdfoutput=1

\usepackage{jheppub}

\usepackage[T1]{fontenc} 
\usepackage{enumitem}

\usepackage{xcolor}
\usepackage{xspace}
\usepackage{amssymb,amsthm,cancel,hyperref,graphicx,xcolor}

\usepackage{subcaption}
\usepackage{comment}

\usepackage{tikz}
\usetikzlibrary{calc}

\usepackage{afterpage}
\usepackage{placeins}

\usepackage{graphicx}
\usepackage{aligned-overset}

\DeclareMathOperator{\De}{d}
\newcommand{\de}{\De\!}
\newcommand{\lc}{\ell_{\text{cut}}}

\newcommand{\RSD}{\bar{R}}
\newcommand{\RSDp}{\bar{R}'}

\def\nn{\nonumber}

\newcommand{\kt}{k_{t}}

\newcommand{\as}{\alpha_s}

\let\originalleft\left
\let\originalright\right
\renewcommand{\left}{\mathopen{}\mathclose\bgroup\originalleft}
\renewcommand{\right}{\aftergroup\egroup\originalright}

\newcommand{\zc}{z_{\rm cut}}

\newcommand{\sherpa}{S\protect\scalebox{0.8}{HERPA}\xspace}
\newcommand{\pythia}{P\protect\scalebox{0.8}{YTHIA}\xspace}
\newcommand{\herwig}{H\protect\scalebox{0.8}{ERWIG}\xspace}

\newcommand{\fastjet}{F\protect\scalebox{0.8}{AST}J\protect\scalebox{0.8}{ET}\xspace}
\newcommand{\rivet}{R\protect\scalebox{0.8}{IVET}\xspace}

\newcommand{\softdrop}{\textsf{SoftDrop}}
\newcommand{\fjcontrib}{\textsf{fjcontrib}}

\newcommand{\zcut}{\ensuremath{z_{\text{cut}}}}

\def\beq{\begin{equation}}  
\def\eeq{\end{equation}}
\def\({\left(}
\def\){\right)}
\def\[{\left[}
\def\]{\right]}

\usepackage{color}
\newcommand{\order}[1]{{\cal O}\left(#1\right)}
\newcommand{\cf}{C_{\text F}}
\newcommand{\ca}{C_{\text A}}

\preprint{INT-PUB-24-049, \, JLAB-THY-24-4244}

\title{\boldmath Heavy Flavour Jet Substructure}

\author[a]{Prasanna K. Dhani,}
\author[b, c, d]{Oleh Fedkevych,} 
\author[e]{Andrea Ghira,} 
\author[e]{Simone Marzani,}
\author[f]{and Gregory Soyez}

\affiliation[a]{Instituto de F\'{i}sica Corpuscular, Universitat de Val\`{e}ncia - Consejo Superior de Investigaciones Cient\'{i}ficas, Parc Cient\'{i}fic, E-46980 Paterna, Valencia, Spain}

\affiliation[b]{Physics and Astronomy Department, Georgia State University, Atlanta, GA 30303, USA}

\affiliation[c]{Center for Frontiers in Nuclear Science, Stony Brook University, Stony Brook, NY 11794, USA}

\affiliation[d]{Jefferson Lab, Newport News, Virginia 23606, USA}

\affiliation[e]{Dipartimento di Fisica, Universit\`a di Genova and INFN, Sezione di Genova,\\ Via Dodecaneso 33, 16146, Genoa, Italy}

\affiliation[f]{Institut de Physique Th\'{e}orique, Paris Saclay University, CNRS, CEA,\\
Orme des Merisiers, B\^{a}t 774, F-91191. Gif-sur-Yvette, France}

\emailAdd{dhani@ific.uv.es}
\emailAdd{ofedkevych@gsu.edu}
\emailAdd{simone.marzani@ge.infn.it}
\emailAdd{andrea.ghira@ge.infn.it}
\emailAdd{gregory.soyez@ipht.fr}
\abstract
{
We present a comprehensive study of energy correlation functions and jet angularities for heavy-flavour QCD jets.
In particular, we discuss the possibility of using these observables to expose the dead cone effect, i.e.\ 
the suppression of collinear QCD radiation around massive quarks, and to investigate the sensitivity of different observable definitions to the presence of quark masses.  
Our calculations are presented as 
all-order resummed predictions at next-to-leading-logarithmic accuracy, matched to (partial) fixed-order results to obtain a better description of the transition around the dead cone threshold.
We also compare our analytic results with \pythia, \herwig and \sherpa Monte Carlo predictions to estimate the impact of non-perturbative contributions such as hadronisation, underlying events and $B$-hadron decays.
}
\begin{document}
\maketitle
\section{Introduction and motivation}\label{sec:intro}
Jets are collimated beams of hadrons abundantly produced in collider experiments. 
The variety of jet production processes makes it possible to use jets to test fundamental properties of the Standard Model (SM)~\cite{Britzger:2017maj, CMS:2013vbb, ATLAS:2017qir, ATLAS:2015yaa, CMS:2014mna, ATLAS:2021qnl, ATLAS:2013pbc, CMS:2014qtp, CMS:2016lna, AbdulKhalek:2020jut, Harland-Lang:2017ytb, Pumplin:2009nk, Watt:2013oha, CMS:2021iwu, ALICE:2021njq} of particle physics as well as for the search for new particles predicted by its extensions~\cite{Soper:2010xk, Godbole:2014cfa, Chen:2014dma, Adams:2015hiv}.
In this context, jet substructure has emerged as a lively field of research~\cite{Marzani:2019hun} at the boundary between theory and experiment and has been employed in a vast number of searches and measurements. 
Furthermore, innovative artificial intelligence techniques are often applied to jet physics, and jet substructure observables are widely used as input to machine learning algorithms~\cite{Larkoski:2017jix,Kasieczka:2019dbj,Benato:2020sbi} and for particle tagging purposes~\cite{Fedkevych:2022mid, Caletti:2021ysv, Dreyer:2021hhr, Cavallini:2021vot, Khosa:2021cyk, Dreyer:2020brq,Baron:2023hkp}.
Moreover, since the formation and evolution of jets are sensitive to the interactions between partons produced via hard scattering process with the dense medium formed in proton-nucleus ($\rm pA$) or nucleus-nucleus ($\rm AA$) collisions~\cite{Lapidus:2017dek,Zapp:2017ria,Tywoniuk:2017dzi,Casalderrey-Solana:2017mjg, Casalderrey-Solana:2019ubu, Mangano:2017plv,Qin:2017roz,Milhano:2017nzm,Chang:2017gkt,KunnawalkamElayavalli:2017hxo, Caucal:2019uvr, Caucal:2020uic,Caucal:2021cfb, Cunqueiro:2023vxl, Chien:2024uax}, jet substructure has become an important tool for studies of the quark-gluon plasma (QGP) properties. 

An accurate description of jets and their structure can be achieved exploiting both fixed-order and resummation techniques in Quantum ChromoDynamics (QCD). However, these calculations are associated with numerous theoretical challenges. 
Despite the fact that an increasing number of processes involving jets are now known to next-to-next-to-leading order (NNLO) accuracy in the strong coupling expansion, the presence of multiple scales, e.g.\ the jet transverse momentum, the jet radius, quark masses, etc., may worsen the convergence of the perturbative expansion due to the presence of large logarithmic contributions at each fixed order, as a consequence of the hierarchy between these scales. While the inclusion of resummation can be beneficial in the case of inclusive observables, i.e.\ jet-radius resummation for inclusive jet cross sections, it becomes mandatory in order to achieve an accurate description of jet substructure. 
However, all-order calculations of jet substructure observables are non-trivial because there are phase-space boundaries that lead to non-global effects~\citep{Dasgupta:2001sh,Dasgupta:2002bw} and because the algorithmic nature of jet definitions makes all-order factorization difficult to achieve.
These challenges have been addressed by the
theory community over the last decade, and a deeper understanding of jet substructure has been achieved thanks to numerous QCD studies (for a review, see e.g.~\cite{Marzani:2019hun} and references therein). 
For instance, the recent calculations of subleading corrections to non-global logarithms~\cite{Banfi:2021xzn,Banfi:2021owj,Becher:2023vrh,Becher:2023mtx,Becher:2023znt} open up the way to perform next-to-next-to-leading logarithmic (NNLL) resummation for jet substructure observables. 
In this context, so-called grooming techniques, i.e.\ algorithms that aim to clean a jet by removing its constituents originating from soft
radiation, significantly improve our ability to use perturbation theory to describe jet physics by reducing the impact of
non-perturbative effects, such as hadronisation and the underlying event (UE) contribution.
In addition, the analytic structure of all-order resummed results is also simplified because groomers, such as the \softdrop~ algorithm~\citep{Larkoski:2014wba}, can eliminate the logarithmic enhancement due to soft gluons at wide angles,
including the intricate structure of non-global logarithms, by converting logarithms of the observable under consideration into logarithms of an external parameter, such as $\zcut$ in the case of \softdrop. Because of these properties, all-order calculations for \softdrop~observables have reached NNLL accuracy and beyond~\cite{Frye:2016aiz,Kardos:2020gty}.
An important and widely-studied class of jet substructure observables are jet angularities~\cite{Larkoski:2014pca, Berger:2003iw, Almeida:2008yp} and  energy-correlation functions (ECFs)~\cite{Larkoski:2013eya,Moult:2016cvt}.
High-precision results for jet angularities  were obtained in refs.~\cite{Caletti:2021oor, Reichelt:2021svh, Kang:2018qra, Kang:2018vgn, Almeida:2014uva, Dasgupta:2022fim, Budhraja:2023rgo}. 
In particular, NLO+NLL$^\prime$ predictions for jet angularities from refs.~\cite{Caletti:2021oor, Reichelt:2021svh} are available as part of 
the resummation plugin to the \sherpa generator framework~\cite{Gerwick:2014gya, Reichelt:2021eru}, which has been used for comparison with experimental measurements~\cite{CMS:2021iwu}.

Recently, precision studies of heavy flavoured jets have gained the attention of the jet substructure community. 
Jets featuring heavy flavours, namely charm ($c$) and beauty ($b$), are of interest for a variety of studies at the Large Hadron Collider (LHC). 
On the one hand, they play a crucial role in studies of the Higgs boson. 
On the other hand, measurements of processes with heavy-flavour jets recoiling against electroweak bosons can be used to probe the heavy-quark component of the proton wave function.
In particular, the recent development of Infra-Red and Collinear (IRC) safe flavour-jet algorithms~\cite{Banfi:2006hf,Caletti:2022hnc,Czakon:2022wam,Gauld:2022lem,Caola:2023wpj} (to NNLO~\cite{Caletti:2022hnc} or to all orders~\cite{Czakon:2022wam,Gauld:2022lem,Caola:2023wpj}) provides us with the possibility of setting up a yet-unexplored flavour-jet substructure program at the LHC.
From a theoretical point of view, calculations for identified heavy flavours can be performed essentially because the quark mass sets a perturbative scale for the running coupling and simultaneously removes collinear singularities. 
From an experimental point of view, the lifetime of $B$ (or $D$) hadrons is long enough for their decay to occur away from the interaction point. Dedicated $b$- and $c$-tagging techniques that exploit this property to identify $B$ and $D$ hadrons or $b$ and $c$ jets are widely used in collider experiments, see e.g.~\cite{ATLAS:2017bcq, ATLAS:2018nnq}.
Resummed calculations for jets initiated by heavy quarks were first performed in the context of studies focusing on $B$-hadron decays~\cite{Aglietti:2006wh,Aglietti:2007bp,Aglietti:2008xn,Aglietti:2022rcm}
and top jets~\cite{Fleming:2007qr,Fleming:2007xt,Bachu:2020nqn,Jain:2008gb,Hoang:2019fze,Bris:2020uyb}, using the formalism of effective field theory.
However, to the best of our knowledge, there exists only a handful of studies addressing jet substructure of heavy-flavour jets~\cite{Maltoni:2016ays, Lee:2019lge,Llorente:2014bha,Li:2017wwc,Li:2021gjw, Craft:2022kdo,Cunqueiro:2022svx,Fedkevych:2022mid,Caletti:2023spr,Blok:2023ugf, Zhang:2023jpe}. 
One of the most interesting QCD effects affecting the substructure of the heavy-flavour jet is the so-called dead-cone effect~\cite{Dokshitzer:1991fd,Dokshitzer:1995ev}, i.e.\ the suppression of collinear radiation around heavy quarks,   
the first direct observation of which was recently reported by the ALICE collaboration~\cite{ALICE:2021aqk}\footnote{For indirect observations of the dead-cone effect, see also refs.~\cite{DELPHI:1992pnf, OPAL:1994cct,  OPAL:1995rqo, SLD:1999cuj, DELPHI:2000edu, ALEPH:2001pfo, ATLAS:2013uet}.}.
The case of heavy-flavour ECFs was considered in ref.~\cite{Lee:2019lge} within the Soft-Collinear Effective Theory (SCET)~\cite{Bauer:2000ew,Bauer:2000yr,Bauer:2001yt,Bauer:2002nz} framework, however from what we know, no studies of the heavy-flavour jet angularities are available in the literature. 
Therefore, in this study we aim to investigate impact of the dead-cone effect on both ECFs and jet angularities for $b$-jets.

This paper is organised as follows: in section~\ref{sec:defs} we consider several possible definitions of ECFs and jet angularities that account for massive constituents. 
Exploiting numerical simulations performed with general-purpose Monte Carlo event generators such as \pythia~\citep{Bierlich:2022pfr,Sjostrand:2006za}, \herwig~\citep{Bellm:2019zci,Corcella:2000bw} and \sherpa~\citep{Sherpa:2019gpd,Gleisberg:2008ta}, we highlight positive and negative features of the different choices. 
In order to put our physics intuition on firmer ground, in section~\ref{sec:FO}, we perform fixed-order calculations at $\mathcal{O}(\as)$ in the quasi-collinear limit~\cite{Catani:2002hc,Catani:2000ef} for the variants of the observables considered in this study.
In section~\ref{sec: resum} we perform analytic resummation and extend the NLL calculations of refs.~\cite{Caletti:2021oor, Reichelt:2021svh} to the case of massive quarks. 
Our calculation resums both logarithms of the observables and of the ratio of the heavy quark mass $m$ to the hard scale of the process.
We consider both ungroomed jets and \softdrop~ones. 
Finally, in section~\ref{sec:MC}, we compare our analytic findings to Monte Carlo simulations, before summarising our findings and conclusions in section~\ref{sec:conclusions}. 
Details of the calculation are provided in the Appendix.
\section{Jet shapes with heavy flavours}\label{sec:defs}

In this section, we perform a rather detailed analysis of possible jet substructure observables that may be employed to study heavy flavour jets. In particular, we focus on  ECFs~\cite{Larkoski:2013eya} and jet angularities \cite{Berger:2003iw,Almeida:2008yp,Larkoski:2014pca} for heavy flavour jets. 

In our considerations we will be guided by simple kinematic arguments, as well as by behaviour of QCD matrix elements in the presence of massive quarks.
In particular, when considering jets initiated by a massive parton, the focus typically shifts from the collinear limit to the quasi-collinear one~\cite{Catani:2002hc,Dhani:2023uxu,Craft:2023aew}. 
Within this approximation, the transverse momentum of the emitted radiation $\kt$ and the mass of the heavy quark are assumed to be small compared to the hard scale of the process. However, their ratio is kept fixed. 
Multiple viable options exist for defining observables, each reducing to the standard definition as the mass approaches zero.
We will exploit the quasi-collinear limit to characterise different definitions of ECFs and angularities.
We will also test the understanding reached with partonic arguments with simulations obtained with general-purpose event generators, focussing on $b$-jets.

\subsection{Definition of the observables}\label{sec:obs}
We begin the discussion by introducing two definitions of ECFs. The first is based on a standard distance in the azimuth-rapidity plane:
\begin{align}
\label{eq:e2_pp}
e_2^\alpha&=  \sum_{i\neq j}\frac{p_{ti} p_{tj}}{p_t^2} \left( \frac{\Delta R_{ij} }{ R_0}\right)^{\alpha}, 
\end{align}
where the sum over $i,j$ runs over constituents of the jet 
clustered with the anti-$\kt$ algorithm~\cite{Cacciari:2008gp} and $p_t$ is the transverse momentum of the jet. 
The distance $\Delta R_{ij}=\sqrt{(y_i-y_j)^2+(\phi_i-\phi_j)^2}$, where $p_{ti}, y_i,\phi_i$ denote the transverse momentum, the rapidity and the azimuthal angle of the $i^{\rm th}$ particle and $R_0$  is the jet radius parameter. The concept of IRC safety requires $\alpha>0$.
The second definition instead uses scalar (dot) products (hence its symbol):
\begin{align}
\label{eq:e2_dot_pp}
\dot{e}_2^\alpha&=  \sum_{i\neq j}\frac{p_{ti} p_{tj}}{p_t^2} \left( \frac{2 p_i \cdot p_j }{p_{ti} R_0\,  p_{tj} R_0}\right)^{\frac{\alpha}{2}}.
\end{align}
We note that, if we neglect the particle masses,  $e_2^\alpha$ and $\dot{e}_2^\alpha$ have the same behaviour in the small $\Delta R_{ij}$ limit. 
However, the ECFs defined as in eq.~(\ref{eq:e2_pp}) and eq.~(\ref{eq:e2_dot_pp}) differ in the case of massive particles. 
To illustrate this point, we take the quasi-collinear limit of eq.~(\ref{eq:e2_dot_pp}), which leads to:
\begin{align}\label{eq:ECF_QC_PP}
\dot{e}_2^\alpha\simeq\sum_{i\neq j} \frac{p_{ti}p_{tj}}{p_t^2} \left(\frac{m_i^2}{p_{ti}^2 R_0^2}+\frac{m_j^2}{p_{tj}^2 R_0^2}+\frac{\Delta R_{ij}^2}{R_0^2}\right)^{\frac{\alpha}{2}}.
\end{align}
Thus, $\dot{e}_2^\alpha$ does not reduce to the standard definition of eq.~(\ref{eq:e2_pp}) in the quasi-collinear limit, due to the presence of the masses $m_i$ and $m_j$. This observable was first studied, in conjunction with \softdrop~in~\cite{Lee:2019lge} using the SCET formalism.

Now let us consider a set of closely related observables generally known as jet angularities. 
The standard jet angularity is defined as 
\begin{equation}
\label{eq:lambda_pp}
\lambda^\alpha=  \sum_{i}\frac{p_{ti} }{p_t} \left( \frac{\Delta R_{i} }{ R_0}\right)^{\alpha},
\end{equation}
where, unlike  eq.~(\ref{eq:e2_pp}), the distance $\Delta R_{i}$ in eq.~(\ref{eq:lambda_pp}) is calculated with respect to the jet axis, which we define by reclustering the jet using the Cambridge/Aachen (C/A) algorithm~\cite{Dokshitzer:1997in,Wobisch:1998wt} with the Winner-Takes-All (WTA) \cite{Larkoski:2014uqa} recombination scheme.
The jet angularites defined as in eq.~(\ref{eq:lambda_pp}) are commonly used for the jets seeded by massless partons, see e.g.~\cite{Reichelt:2021svh, Caletti:2021oor}.
In the case of heavy jets, however, it is interesting to consider different definitions of the reference axis in eq.~(\ref{eq:lambda_pp}) to highlight the effects due to the quark masses.
In order to do that we introduce two different four-vectors: $n_0$ is
a massless four-vector with the same rapidity $y$ and azimuth $\phi$
of the WTA axis, and unit transverse momentum, while in $n$ we keep
the mass ($m_n$) of the particle that is aligned with the WTA axis:
\begin{subequations}
\begin{align}
n_0=&\left(\cosh y, \cos\phi,\sin\phi,\sinh y\right),\\
 n=& \left(\frac{m_{t}}{p_{t}}\cosh y, \cos\phi,\sin\phi,\frac{m_{t}}{p_{t}}\sinh y\right),
\end{align}
\end{subequations}
where $m_{t}^2= p_{t}^2+m_n^2$.
With these considerations in mind, we define four additional variants of jet angularities:
\begin{subequations}
\begin{align}
\label{eq: lambda dot pp}
\dot{\lambda}_0^\alpha=  \sum_{i}\frac{p_{ti} }{p_t} \left( \frac{2 p_i \cdot n_0}{p_{ti} R_0^2}\right)^\frac{\alpha}{2}, \qquad& \dot{\lambda}^\alpha=  \sum_{i}\frac{p_{ti} }{p_t} \left( \frac{2 p_i \cdot n}{p_{ti} R_0^2}\right)^\frac{\alpha}{2},
 \\
\label{eq: lambda circle pp}
\mathring{\lambda}_0^\alpha=  \sum_{i\neq n}\frac{p_{ti} }{p_t} \left( \frac{2 p_i \cdot n_0}{p_{ti} R_0^2}\right)^\frac{\alpha}{2}, \qquad&
\mathring{\lambda}^\alpha=  \sum_{i\neq n}\frac{p_{ti} }{p_t} \left( \frac{2 p_i \cdot n}{p_{ti} R_0^2}\right)^\frac{\alpha}{2},
\end{align}
\end{subequations}
where the sum $i$ is over the jet's constituents. 
We note that while  eq.~(\ref{eq:lambda_pp}) only made use of the transverse momentum fraction of $i^{\rm th}$ particle and its distance from the WTA axis in the azimuth-rapidity plane, the definitions in eqs.~(\ref{eq: lambda dot pp}) and~(\ref{eq: lambda circle pp}) instead consider the scalar products between the four momenta of the individual particles and the two different four-vectors $n$ and $n_0$ defined above.
In particular, in the last two cases, denoted by open circles, we exclude the contribution of the particle aligned with the WTA axis. As far as massless particles (and axis) are concerned, all five definitions of angularities in the collinear limit behave in the same way. If instead we consider massive particles and perform the quasi-collinear limit, we find that
\begin{subequations}
\begin{align}
\label{eq: lambda dot pp approx}
\dot{\lambda}_0^\alpha &\simeq  \sum_{i \neq n}\frac{p_{ti} }{p_t} \left(\frac{m_i^2}{p_{ti}^2 R_0^2}+\frac{\Delta R_i^2}{R_0^2}\right)^\frac{\alpha}{2}+ \left(\frac{m_n^2}{p_{t}^2 R_0^2}\right)^{\frac{\alpha}{2}}, \,\, \nonumber \\\dot{\lambda}^\alpha &\simeq \sum_{i \neq n}\frac{p_{ti} }{p_t} \left(\frac{m_i^2}{p_{ti}^2 R_0^2}+\frac{m_n^2}{p_{t}^2 R_0^2}+\frac{\Delta R_i^2}{R_0^2}\right)^\frac{\alpha}{2}+ \left(\frac{2 m_n^2}{p_{t}^2 R_0^2}\right)^{\frac{\alpha}{2}},
\\
\label{eq: lambda circle pp approx}
\mathring{\lambda}_0^\alpha &\simeq  \sum_{i \neq n}\frac{p_{ti} }{p_t} \left(\frac{m_i^2}{p_{ti}^2 R_0^2}+\frac{\Delta R_i^2}{R_0^2}\right)^\frac{\alpha}{2}, \qquad\qquad\,
\mathring{\lambda}^\alpha \simeq \sum_{i \neq n}\frac{p_{ti} }{p_t} \left(\frac{m_i^2}{p_{ti}^2 R_0^2}+\frac{m_n^2}{p_{t}^2 R_0^2}+\frac{\Delta R_i^2}{R_0^2}\right)^\frac{\alpha}{2}.
\end{align}
\end{subequations}
Thus, we note that both $\dot{\lambda}_0^\alpha$ and $\dot{\lambda}^\alpha$ do not vanish in the soft and collinear limit, but rather saturate to a constant proportional to the mass of the particle that is aligned to the WTA axis. On the other hand, if we consider $\mathring{\lambda}_0^\alpha$ and $\mathring{\lambda}^\alpha$, for which the particle aligned to the WTA axis does not enter in the sum, this contribution is absent.

We also compute groomed versions of the above observables.
They are defined by first applying \softdrop~\cite{Larkoski:2014wba} to the jets and then use the groomed jet's constituents to compute the observables.
Note that we always keep $p_t$ to be the transverse momentum of the ungroomed jet. 
\softdrop~recursively removes soft-wide angle constituents from a jet. Let us briefly recall this procedure. The algorithm first  re-clusters a given jet with radius $R_0$ and transverse momentum $p_t$ with the C/A algorithm. It then parses the resulting angular-ordered branching history, grooming away the softer branch, until the condition 
\begin{align}
    \label{eq:sdcrit}
\frac{\min \left( p_{t1}, p_{t2}\right)}{p_{t1} + p_{t2}}  &> \zc\left( \frac{\Delta R_{12}}{R_0}\right)^\beta
\end{align}
is satisfied. 
In the expression above, $1$ and $2$ denote the branches at a given step in the clustering, $p_{t1}$ and $p_{t2}$ are the corresponding transverse momenta, and $\Delta R_{12}$ is their rapidity-azimuth separation.  
In what follows, we  perform resummed calculations that apply for generic values of $\beta \ge0$ and $\zc$. However, when presenting fixed-order and numerical results, we  focus on $\beta=0$, which corresponds to the modified-Mass-Drop-Tagger (mMDT)~\cite{Dasgupta:2013ihk}. The use of mMDT is particularly motivated in our study because we want to extract information about the modification of the behaviour of jet substructure observables because of the dead cone. Therefore, we want to concentrate on collinear dynamics and mMDT efficiently removes contributions from soft emissions. 
However, we should point out that the use of mMDT may lead to some peculiar behaviour in the $\dot{e}_2^\alpha$ and $\mathring{\lambda}^\alpha$ distributions, which is reminiscent of what happens for $\dot{\lambda}^\alpha_0$ and $\dot{\lambda}^\alpha$. This is because the mMDT condition imposes a lower bound for $p_{ti}$. Because of this additional constraint, eq.~(\ref{eq:ECF_QC_PP}) and the second equation of~(\ref{eq: lambda circle pp approx}) do not vanish in the collinear limit, but rather saturate to a constant proportional to $\zc$ times mass of the particle that is aligned with the WTA axis, to some power. We discuss the numerical size of this effect in the next section.
\subsection{First Monte Carlo studies}\label{sec:MC_part1}
\begin{figure}
 \includegraphics[width=0.5\textwidth,page=2]{figures/dot-variants.pdf}
 \includegraphics[width=0.5\textwidth, page=1]{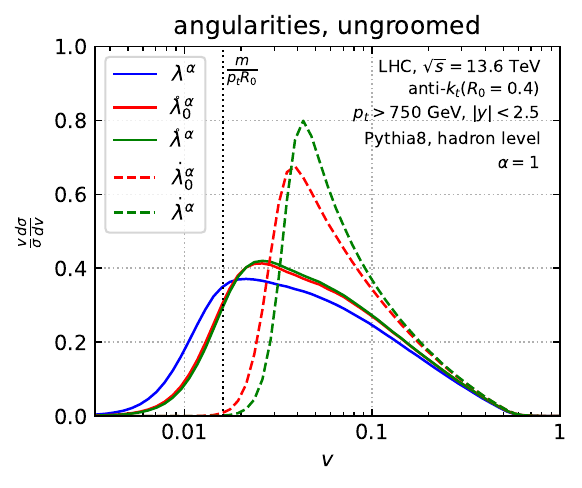}
  \caption{Distributions of all variants introduced in section~\ref{sec:obs}, with $\alpha=1$, measured on $b$-tagged jets. No grooming has been applied. The left plot is for the ECFs, the right plot for the angularities.
  }\label{fig:MC-distr-ungroomed}
\end{figure}

\begin{figure}
 \includegraphics[width=0.5\textwidth,page=4]{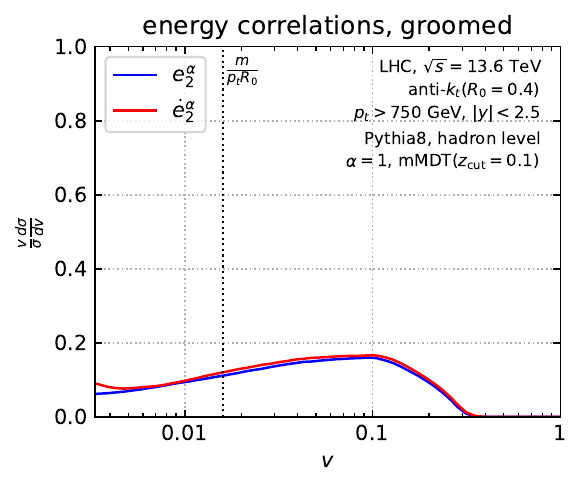}
 \includegraphics[width=0.5\textwidth, page=3]{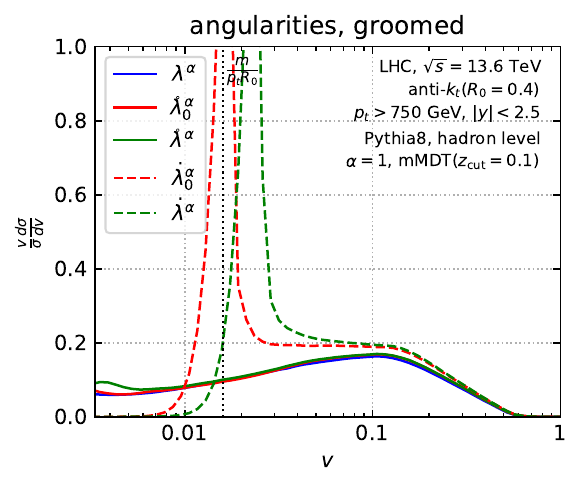}
  \caption{Same observables as in figure~\ref{fig:MC-distr-ungroomed}, but for $b$-jets groomed with mMDT, with $\zc=0.1$.}\label{fig:MC-distr-groomed}
\end{figure}

We analyse the observables defined above exploiting Monte Carlo simulations. To this purpose, we generate samples of heavy- and light-jet events produced via proton-proton ($pp$) collisions.
Since the main goal of this work is a theoretical study of heavy-quark emission patterns, we consider a simplified setup where the heavy-jet samples are generated only via  $pp \rightarrow b\bar{b}$ processes (as a consequence for the light-jet sample we consider production of $u\bar{u}$, $d\bar{d}$ and $s\bar{s}$ pairs). 
Such a setup is sufficient to study the suppression of collinear radiation due to the $b$-quark mass, but of course more sophisticated simulations and flavour-labelling procedures are required to make realistic phenomenological predictions, especially in view of the $g \to b \bar{b}$ splittings that increasingly happen in high $p_t$ jets.

Our Monte Carlo simulations are performed with the general-purpose event generators \pythia, \herwig and \sherpa, as each of these programs implements heavy particle radiation in its own way, see refs.~\cite{Norrbin:2000uu, Bierlich:2022pfr}, \cite{Gieseke:2003rz, Hoang:2018zrp, Cormier:2018tog} and \cite{Schumann:2007mg,Krauss:2016orf} respectively. The generated events are analysed using the \rivet framework~\cite{Buckley:2010ar,Bierlich:2019rhm}. 
In this section, we only show representative results obtained with \pythia, while a comparison between the different generators will be performed in section~\ref{sec:MC}.
The particles in the final state are clustered into jets using the
anti-$k_t$ jet algorithm~\cite{Cacciari:2008gp} with jet radius parameter $R_0 = 0.4$ and the standard $E$-scheme recombination scheme as implemented in the  \fastjet code~\cite{Cacciari:2011ma}.
As anticipated, we use the \softdrop~algorithm for grooming with the parameters $\zc = 0.1$ and
$\beta = 0$~\cite{Marzani:2017mva, Larkoski:2014wba}, i.e.\ mMDT, and the C/A algorithm \cite{Dokshitzer:1997in, Wobisch:1998wt}
for jet reclustering, as implemented in the \fjcontrib~library.
We select central jets with rapidities $|y|<2.5$ and a fairly high transverse momentum \mbox{$p_t>750$~GeV}. This may not be the best choice for phenomenological studies of $b$-jets, but we wanted to keep the hard scale $p_t R_0$ large enough so that non-perturbative corrections are under control.

We first consider the two variants of the ECFs, focusing on the case $\alpha=1$. The distributions for the ungroomed case are shown in figure~\ref{fig:MC-distr-ungroomed}, left. The distributions for $e^\alpha_2$ and $\dot{e}^\alpha_2$ are shown in blue and red respectively. The vertical dotted line always represents the value of the observable at which we probe the dead cone, as determined by a calculation with a single gluon emission. For $\alpha=1$ this is $v=\frac{m}{p_t R_0}$, where $m$ denotes the mass of the heavy quark. The two distributions have a similar shape, with the $\dot{e}^\alpha_2$ shifted to right, as one might expect from eq.~(\ref{eq:ECF_QC_PP}). 
While Monte Carlo simulations show no significant difference between the two variants of the ECFs, the opposite is the case when we analyse the jet angularities. As with the ECFs, we focus on $\alpha=1$. The results are shown in figure~\ref{fig:MC-distr-ungroomed} on the right.
We start by considering $\lambda^\alpha$, eq.~(\ref{eq:lambda_pp}). The corresponding distribution is shown in blue.
The result is rather similar to the corresponding case for the ECFs, namely $e_2^\alpha$. This is expected as both definitions are purely based on $\Delta R$ and transverse-momentum fractions and have the same soft and quasi-collinear limit.

Next, we consider the variants of angularity defined via scalar products. We start with $\dot{\lambda}^\alpha_0$ and $\dot{\lambda}^\alpha$, which are given by a sum over all particles in the jets, including the particle that is aligned with the WTA axis, either massless or massive, see eq.~(\ref{eq: lambda dot pp}). We have already noticed that these observables do not vanish in the soft and quasi-collinear limit. Indeed, the corresponding distributions, shown in dashed red lines ($\dot{\lambda}_0^\alpha$) and dashed green lines ($\dot{\lambda}^\alpha$), exhibit an endpoint behaviour at low values of the observables.
The end-point is close to $\frac{m}{p_t R_0}$, which is expected because, in the soft and collinear region, the $b$-quark is likely to be aligned with the WTA axis. 
Note that the minimum value of $\dot{\lambda}^\alpha$ is larger than that of  $\dot{\lambda}^\alpha_0$, which is in agreement with eq.~(\ref{eq: lambda dot pp approx}). 
The radiation of soft and collinear gluons mostly contributes just above the end points, leading to very visible spikes in the distributions.
We argue that these observables should be avoided for the following reason. Our aim is to investigate the dynamical suppression of collinear radiation, i.e.\ the dead-cone. The definitions of the observables $\dot{\lambda}^\alpha_0$ and $\dot{\lambda}^\alpha$  effectively introduce a kinematic, mass-dependent cut-off for the distribution. This effect is larger than that of the dead-cone and completely overshadows it. Therefore, we  discard $\dot{\lambda}^\alpha_0$ and $\dot{\lambda}^\alpha$ from our analytic studies.
The angularity variants $\mathring{\lambda}^\alpha_0$ and $\mathring{\lambda}^\alpha$ behave better than their dotted counterparts precisely because we avoid the particle aligned with the WTA axis in their definitions, see eq.~(\ref{eq: lambda circle pp}). The distributions for these observables are shown in figure~\ref{fig:MC-distr-ungroomed}, right, with solid red and green curves. The two distributions are very close to each other and rather similar to $\dot{e}_2^\alpha$.
This is not entirely expected since, according to eq.~(\ref{eq: lambda circle pp approx}), in the presence of massless soft and collinear radiation, the behaviour of $\mathring{\lambda}_0^\alpha$ is more similar to $\lambda^\alpha$ than $\mathring{\lambda}^\alpha$. We investigate this feature in the next section, where we perform a fixed-order study of these distributions.

We now analyse the groomed counterparts of all the observables discussed so far. The results are shown in figure~\ref{fig:MC-distr-groomed}, on the left for the ECFs and on the right for the angularities. 
We first note that the spikes for the pathological variants $\dot{\lambda}^\alpha_0$ and $\dot{\lambda}^\alpha$ are even more pronounced in the case of mMDT jets.
Moreover, we see very little difference between the two variants of ECFs and between the other angularities, except for the very small observable range. As anticipated in the previous section, we find that both $\dot{e}_2^\alpha$ and $\mathring{\lambda}^\alpha$  start to exhibit some features that we believe originate from the endpoint of the distribution (not visible on this plot). Due to $\zc$ suppression, this affects the distribution at much lower values of the observable. 

\subsection{The role of $B$-hadron decay}\label{sec:b-decays}
\begin{figure}
\includegraphics[width=0.33\textwidth,page=3]{figures/bdecays.pdf}
\includegraphics[width=0.33\textwidth,page=1]{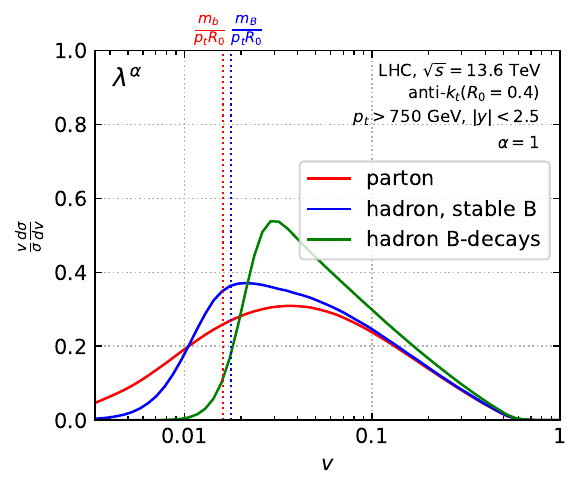}
\includegraphics[width=0.33\textwidth,page=2]{figures/bdecays.pdf}
\includegraphics[width=0.33\textwidth,page=6]{figures/bdecays.pdf}
\includegraphics[width=0.33\textwidth,page=4]{figures/bdecays.pdf}
\includegraphics[width=0.33\textwidth,page=5]{figures/bdecays.pdf}
  \caption{Effect of the decays of the $B$-hadron on a representative selection of observables, from left to right the ECF $e_2^\alpha$, and the jet angularities $\lambda^\alpha$ and $\dot{\lambda}^\alpha_0$, all for $\alpha=1$. The top plots are for ungroomed jets, the bottom ones for mMDT jets. }\label{fig:bdecays}
\end{figure}
The simulations presented in the previous section were obtained keeping the produced $B$ hadrons stable and hence using their four momenta as input to the jet shapes. This is of course not realistic, as $B$ hadrons decay inside the detector, giving rise to displaced vertices, which is one of the main features exploited by $b$-tagging algorithms. It has already been shown in ref.~\cite{Lee:2019lge} that using the decay products of the $B$ hadron to build up the jet shape can strongly affect the shape of the distribution. Since this study was performed for the groomed version of $\dot{e}_2^\alpha$, we decided to repeat it here for a wider class of observables, before and after application of jet grooming. 
Besides $e_2^\alpha$, $\lambda^\alpha$,  as a representative  we also choose  one of the angularities with strong sensitivity to the heavy-flavour mass, namely $\dot{\lambda}^\alpha_0$. In figure~\ref{fig:bdecays} we show the distributions for these three observables, measured on ungroomed jets (top) and mMDT jets (bottom). The details of the simulation are the same as in the previous section. We show the results for three different simulation levels:  parton level, i.e.\ after the parton shower, in red, hadron level with stable $B$ hadron in blue, with $B$ decay in green. As before, we consider the $\alpha=1$ case.

All distributions are affected by $B$ decays in a rather significant way. The distortion appears to be stronger when jet angularities with grooming are considered. However, we note that the ECF $e_2^\alpha$, in the vicinity of the dead-cone region, appears to be less sensitive to this effect, both in the ungroomed and groomed case. In any case, the general picture that emerge is that, while hadronisation corrections appear to be under control, at least for groomed jets (see also section~\ref{sec:MC}), $B$ decays can ruin this picture. Thus, we fully agree with the conclusions of ref.~\cite{Lee:2019lge} and encourage measurements of heavy-flavour jet substructure observables that as an input use the reconstructed momentum of the heavy-flavour hadron rather than the momenta of its decay products. 
This is usually achieved through the experiments by reconstructing the kinematics of a specific decay channel, see e.g.\ \cite{Bainbridge:2020pgi, CMS:2017ygm, ATLAS:2021agf, ALICE:2021aqk, ALICE:2022tji} for studies on $B$ and $D$ mesons. With this information at hand, the observables discussed in this study can be then measured using the momentum of the reconstructed $B(D)$ hadron. 
However, measurements that exploit the kinematics of a specific decay channel may be limited by statistics. This issue can be avoided by considering $b$-($c$-) jets that have been tagged inclusively. 
The situation is here more delicate, from experimental point of view because one does not reconstruct the full kinematics of the decay. However, very recently, the CMS collaboration has developed an innovative machine-learning strategy to aggregate particles from the decay vertex into a pseudo $B$ hadron in $b$-tagged jets,~\cite{CMS-PAS-HIN-24-005} and ATLAS is also working on similar studies~\footnote{We thank Leticia Cunqueiro Mendez (CMS) and Federico Sforza (ATLAS) for discussions on this issue.}. An assessment of the uncertainty related to this procedure is an important component of the experimental error budget of on-going and future measurements. On the theory side, a complete phenomenological study of these observables, which will plan to do in the near future, will have to consider the hadronisation correction that maps our parton-level result to the unfolded objects provided by the experiments.
\section{Fixed-order analysis}\label{sec:FO}
In this section, we study the observables defined above  in perturbation theory. 
In particular, we are going to compute the cumulative distribution for the ECFs and the jet angularities, with the exception of  $\dot{\lambda}^\alpha_0$ and $\dot{\lambda}^\alpha$, which we consider pathological. We therefore define
\begin{equation}
\Sigma_V(v) = \frac{1}{\sigma_0}\int_0^v \de v' \frac{\de \sigma_V}{\de v'},
\end{equation}
i.e.~the probability for the observable $V$ to be smaller than some given value $v$. The subscript $V$ reminds us of the fact that there can be differences between the cumulative distributions of the various observables. Indeed, one of the objectives of this section is to study at what logarithmic accuracy these differences occur and to assess their numerical relevance.
We consider the $\mathcal{O}(\as)$ contribution to $\Sigma_V(v)$, where $\as$ is the strong coupling. This corresponds to the one-gluon emission, plus $\mathcal{O}(\as)$ virtual corrections:
\begin{equation}
\Sigma_V(v)=1+ \Sigma_V^{(1)} (v)+\dots
\end{equation}
QCD scattering amplitudes with massive partons factorise in the quasi-collinear limit.
As already stated, in this approximation the transverse momentum of the emitted radiation $\kt$ and the mass of the heavy quark are assumed to be small compared to the hard scale of the process. However, their ratio is kept fixed. 
Using the standard Sudakov decomposition, we obtain in this approximation
\begin{equation}\label{eq:QC-fact}
|\mathcal{M}|^2\simeq \frac{8\pi \as z(1-z)}{\kt^2+z^2 m^2} P_{gb}(z,\kt^2) |\mathcal{M}_0|^2,
\end{equation}
where $\mathcal{M}$ denotes the original scattering amplitude and $\mathcal{M}_0$ denotes the reduced amplitude with one fewer external parton. In eq.~(\ref{eq:QC-fact}), $z$ denotes the fraction of the momentum transferred in the splitting process $b \rightarrow g + b$, and $P_{gb}$ is the leading order time-like massive splitting function:
\begin{equation}
P_{gb}(z,\kt^2)= \cf \left(\frac{1+(1-z)^2}{z}-\frac{2 m^2 z(1-z)}{\kt^2+z^2 m^2}\right).
\end{equation}

In order to compute the $\mathcal{O}(\as)$ contribution $ \Sigma_V^{(1)}$ in the quasi-collinear limit, we integrate eq.~(\ref{eq:QC-fact}) together with a $\Theta$ function that forces the observable to be smaller than a given value $v$. In the language of resummation, this corresponds to the first-order approximation of (minus) the radiator, $-\mathcal{R}^{(\text{f.o.})}_V$.  Thus, we have to evaluate:
 \begin{align}\label{eq: Sudakov fo}
\mathcal{R}^{(\text{f.o.})}_V(v,\xi)&= -\frac{\as}{2\pi}  \int^1_0 \de z\int^{Q^2}_0 \frac{\de \kt^2}{\kt^2+z^2 m^2}\,P_{gb}(z,\kt^2)\,\Big[ \Theta(v-V(\kt,\eta)) -1\Big] \nonumber\\
&=\frac{\as}{2\pi} \int^1_0 \de z\int^{Q^2}_0 \frac{\de \kt^2}{\kt^2+z^2 m^2}\, P_{gb}(z,\kt^2)\,  \Theta(V(\kt,\eta)-v),
 \end{align}
where  $\eta$ is the rapidity of the emission with respect to the emitting
particle and $Q=p_t R_0$ is the hard scale of the process.
 The $-1$ contribution accounts for virtual corrections, while $V(\kt,\eta)$ parameterises the particular observable we consider in the quasi-collinear limit.
When performing the above integrals, we  neglect power-corrections both in the observable $v$ and in the ratio of the heavy-quark mass to the hard scale of the process, but we keep the full dependence on their ratio
\begin{equation}\label{eq:x-ratio}
x=\frac{\xi}{v^\frac{2}{\alpha}}, \quad \text{with}  \quad \xi=\frac{m^2}{Q^2}.
\end{equation}

We find it easier to begin with the angularities, for which we can straightforwardly arrive at an analytic expression. 
First, let us note that if we consider $\lambda^\alpha$ and $\mathring{\lambda}_0^\alpha$ in the quasi-collinear limit, eq.~(\ref{eq: lambda circle pp approx}), with just one gluon emission, we immediately see that $\lambda^\alpha\simeq \mathring{\lambda}_0^\alpha$. Substituting the expression for the observable in the quasi-collinear limit, $V=\frac{\kt^\alpha z^{1-\alpha}}{Q^\alpha}$, in eq.~(\ref{eq: Sudakov fo}), we find
\begin{align}\label{eq: Sudakov for lambda a}
  \mathcal{R}^{(\text{f.o.})}_{\lambda^\alpha}(v,\xi)
  &=\mathcal{R}^{(\text{f.o.})}_{\mathring{\lambda}_0^\alpha}(v,\xi)\\
  &= \frac{\as\cf}{\pi}\Bigg[
    \frac{1}{\alpha}\log^2 v
    -\frac{\alpha}{4}\log^2(1+x)
    +\frac{3}{2\alpha}\log v
    +\left(\frac{3}{4}-\frac{\alpha}{2}\right)\log(1+x)
    +\frac{7}{4\alpha}\nonumber \\
&+\frac{\alpha-2}{\alpha+2}x\,_2 F_1\left(1,1+\frac{\alpha}{2};2+\frac{\alpha}{2};-x\right)+\frac{x}{4(\alpha+1)}\,_2 F_1\left(1,1+\alpha;2+\alpha;-x\right)-\frac{\alpha}{2}\text{Li}_2\left(\frac{x}{1+x}\right)\Bigg]. \nonumber
\end{align}
Similarly, the cumulative distribution for $\mathring{\lambda}^\alpha$, for which $V=z^{1-\alpha}\left( \frac{\kt^2}{Q^2}+ z^2 \xi\right)^\frac{\alpha}{2}$, is found to be
\begin{subequations}
\begin{align}\label{eq: Sudakov for circ lambda a x<1}
  \mathcal{R}^{(\text{f.o.})}_{\mathring{\lambda}^\alpha}(v,\xi)
  \overset{x<1}&{=} \frac{\as\cf}{\pi}\left(
    \frac{1}{\alpha}\log^2 v +\frac{3}{2\alpha}\log v
    +\frac{7}{4\alpha}-\frac{x\alpha^2}{2(\alpha+2)}
    \right); \\ \label{eq: Sudakov for circ lambda a x>1}
  \overset{x>1}&{=} \frac{\as\cf}{\pi}\bigg[
    \frac{1}{\alpha}\log^2 v
    -\frac{\alpha}{4}\log^2 x
     +\frac{3}{2\alpha }\log v
    +\left(\frac{3}{4}-\frac{\alpha}{2}\right)\log x
   \nn\\
    &\phantom{\overset{x>1}{=} \frac{\as\cf}{\pi}\bigg[}+1-\frac{\alpha}{2}
    +\frac{16-(2+\alpha)x^{-\frac{\alpha}{2}}}{4\alpha(2+\alpha)}x^{-\frac{\alpha}{2}}\bigg].
\end{align}
\end{subequations}
We notice that the expression found for $\mathring{\lambda}^\alpha$ is much simpler than that for $\lambda^\alpha$ and $\mathring{\lambda}^\alpha_0$. 
The result appears as the sum of two regions, each containing only one elementary function. 
This is in contrast to eq.~(\ref{eq: Sudakov for lambda a}), where we find hypergeometric functions. This is due to the fact that the scalar product that appear in the $\Theta$ function for $\mathring{\lambda}^\alpha$ has the same structure as the denominators that appear in the matrix elements squared taken in the quasi-collinear limit. Thus, the transverse-momentum integral in eq.~(\ref{eq: Sudakov fo}) can be simplified by shifting the integration variable. However, this simplification does not occur for $\lambda^\alpha$ and $\mathring{\lambda}^\alpha_0$.

We now examine behaviour of the above expressions at small and large $x$ values. The former is the limit in which the quark mass is much smaller than the observable, and hence we expect to recover the massless result for the cumulative distribution. In the opposite limit, we expect to probe the dead cone region instead. In the case of $\lambda^\alpha$ we find at small $x$:
\begin{subequations} \label{eq:lambda-x}
  \begin{align} \label{eq:lambda-smallx}
    \mathcal{R}^{(\text{f.o.})}_{\lambda^\alpha}(v,\xi)\overset{x\ll 1}{=}\frac{\as \cf}{\pi}\left(\frac{1}{\alpha}\log ^2 v+\frac{3}{2 \alpha}\log v+ \frac{7}{4 \alpha}+ \order x\right), 
  \end{align}
while at large $x$: 
\begin{align}\label{eq:lambda-largex}
  \mathcal{R}^{(\text{f.o.})}_{\lambda^\alpha}(v,\xi)
  \overset{x\gg 1}&{=}\frac{\as \cf}{\pi} \Bigg[\frac{1}{\alpha}\log ^2 v-\frac{\alpha}{4}\log^2 x+\frac{3}{2 \alpha}\log v+ \left(\frac{3}{4}-\frac{\alpha}{2}\right)\log{x}
 -\frac{\alpha\pi^2}{12}+1 +\order{x^{-\frac{\alpha}{2}}}\Bigg]
\nn\\
&=\frac{\as \cf}{\pi} \Bigg[\log \xi\, \log v-\frac{\alpha}{4}\log^2 \xi+ \left(\frac{3}{4}-\frac{\alpha}{2}\right)\log{\xi}+\log v-\frac{\alpha\pi^2}{12}+1 +\order{x^{-\frac{\alpha}{2}}}\Bigg].
  \end{align}
\end{subequations} 
We note that eq.~(\ref{eq:lambda-smallx}) exhibits a double logarithmic behaviour due to soft and collinear emissions. Moreover, the coefficient of the single logarithm, which is of collinear origin, is proportional to $B_1=-\frac{3}{2}\cf$ as expected. In eq.~(\ref{eq:lambda-largex}) instead, double logarithms of $v$ between the first and second contributions cancel out, leaving behind double logarithms of $\xi$. In fact, the single collinear logarithm is also replaced by a logarithm of the mass, see e.g.\ the discussion in~\cite{Gaggero:2022hmv,Ghira:2023bxr}. 
Finally, let us note that the finite contributions in the two limits are also different.

\begin{figure}
 \includegraphics[width=0.55\textwidth,page=1]{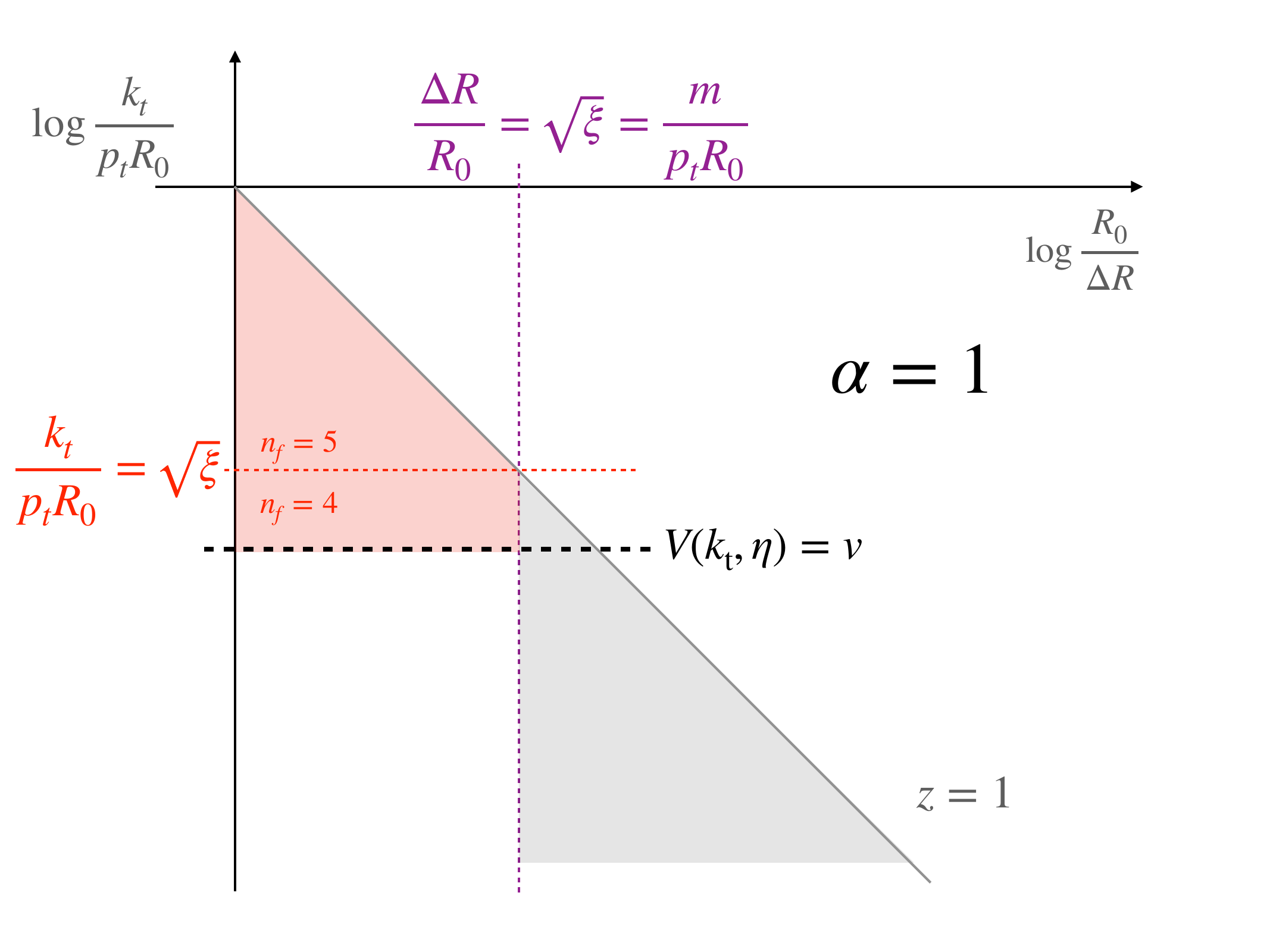} 
 \hspace{-1cm}
 \includegraphics[width=0.55\textwidth, page=2]{figures/lund.pdf}
  \caption{Lund plane representation of the soft and quasi-collinear phase space for emissions off a $b$ quark.  On the left-hand panel we show the case of an ungroomed jet,  while on the right-hand plane, the case of a groomed jet. The dead-cone region is indicated in grey. When \softdrop~is applied,  the groomed-away region appears in light purple. Where we show the case of mMDT, i.e.\ \softdrop~ with $\beta=0$. The dashed line in black indicates a measurement of one of the jet shapes considered in this paper, either ECF or jet angularity. We explicitly show the case of $\alpha=1$.  The corresponding area in pink is the vetoed region, which gives rise to the Sudakov form factor. Finally, horizontal dotted red lines indicate the boundaries between different flavour numbers for the running coupling.}\label{fig:lund-plane}
\end{figure}
 
It is useful to discuss these results in the form of Lund diagrams~\cite{Andersson:1988gp, Dreyer:2018nbf}. The primary Lund plane for a jet originated by a massive quark is depicted in the figure~\ref{fig:lund-plane} on the left.
On the horizontal axis, we have the logarithm of the distance $\Delta R$ in units of the jet radius $R_0$, while on the vertical axis, we have the transverse momentum of the emission with respect to the massive quark, normalised to the hard scale. The emissions at angles larger than the angle of the dead cone $\Delta R=\frac{m}{p_t}$ do not give rise to a logarithmically enhanced contribution, and this region, i.e.\ the dead cone, is shaded in grey in the Lund plane. In the soft and quasi-collinear limits, all shape variables considered in this work are represented by a straight line (dashed black) whose slope is determined by the angular \mbox{exponent $\alpha$}. In figure~\ref{fig:lund-plane}, the case $\alpha=1$ is depicted. The area in red corresponds to the radiator. As one can see, the presence of the dead-cone (shown in dotted purple) induces two different regions. Above $\xi^\frac{\alpha}{2}$ the radiator does not feel the presence of the dead cone effect, and has the same form as for a massless quark, i.e.\ double logarithms of the observable. This corresponds to the $x \to 0$ limit discussed above. Below $\xi^\frac{\alpha}{2}$ the radiator is instead sensitive to the dead-cone boundary, which cuts off the collinear region, leaving only single logarithms of $v$ but double logarithms of the mass. This corresponds to the $x \to \infty$ limit above.

Next, we consider  small- and large-$x$ limits for the cumulative distribution of the angularity $\mathring{\lambda}^\alpha$ (eqs.~(\ref{eq: Sudakov for circ lambda a x<1}) and  (\ref{eq: Sudakov for circ lambda a x>1})) and we find
\begin{subequations} \label{eq:lambda-circle-x}
  \begin{align}\label{eq:lambda-circle-smallx}
    \mathcal{R}^{(\text{f.o.})}_{\mathring{\lambda}^\alpha}(v,\xi)
    \overset{x\ll 1}&{=} \frac{\as \cf}{\pi}\left(\frac{1}{\alpha}\log ^2 v+\frac{3}{2 \alpha}\log v+ \frac{7}{4 \alpha}+ \order x\right);\\
 \label{eq:lambda-circle-largex} 
    \mathcal{R}^{(\text{f.o.})}_{\mathring{\lambda}^\alpha}(v,\xi)
    \overset{x\gg 1}&{=} \frac{\as \cf}{\pi} \Bigg[\frac{1}{\alpha}\log ^2 v-\frac{\alpha}{4}\log^2 x+\frac{3}{2 \alpha}\log v+ \left(\frac{3}{4}-\frac{\alpha}{2}\right)\log{x}
 -\frac{\alpha}{2}+1+\order{x^{-\frac{\alpha}{2}}}\Bigg]
\nn\\
    &=\frac{\as\cf}{\pi} \Bigg[
      \log \xi\, \log v-\frac{\alpha}{4}\log^2 \xi+ \left(\frac{3}{4}-\frac{\alpha}{2}\right)\log{\xi}+\log v-\frac{\alpha}{2}+1 +\order{x^{-\frac{\alpha}{2}}}\Bigg].
 \end{align}
 \end{subequations}
As expected, the massless limit, eq.~(\ref{eq:lambda-circle-smallx}), coincides  with the result for $\lambda^\alpha$,  see  eq.~(\ref{eq:lambda-smallx}). In the opposite limit, i.e.\ large $x$, all logarithmic contributions are the same as in eq.~(\ref{eq:lambda-largex}), but with a different constant term. This suggests that $\lambda^\alpha$ and $\mathring{\lambda}^\alpha$, in the large $x$ region, are the same up to next-to-leading logarithmic (NLL) accuracy, but start to differ at next-to-NLL (NNLL) level and beyond.

The calculation for the ECFs $e_2^\alpha$ and $\dot{e}^\alpha_2$ in the quasi-collinear limit, see eqs.~(\ref{eq:e2_pp}) and (\ref{eq:ECF_QC_PP}), is more cumbersome because of the products $p_{ti}\,p_{tj}$, which did not allow us to obtain a closed-form analytic expression for generic $\alpha$. 
For $e_2^\alpha$, we use $V=\frac{\kt^\alpha z^{1-\alpha}(1-z)}{Q^\alpha}$, and we find
\begin{align}\label{eq: Sudakov for ECF}
\mathcal{R}^{(\text{f.o.})}_{e_2^{\alpha}}(v,\xi)=\mathcal{R}^{(\text{f.o.})}_{\lambda^\alpha}(v,\xi)
+\frac{\as \cf}{\pi}\Bigg\{\int^1_0 \de z \frac{1+(1-z)^2}{2z}\log\left(\frac{(1-z)^\frac{2}{\alpha}+((1-z)z)^\frac{2}{\alpha}x}{1+((1-z)z)^\frac{2}{\alpha}x}\right)
\nn\\
+\int^1_0 \de z~
\frac{1-z}{z}\left[\frac{z^{\frac{2}{\alpha}}x}{1+z^\frac{2}{\alpha}x}-\frac{((1-z)z)^{\frac{2}{\alpha}}x}{1+((1-z)z)^\frac{2}{\alpha}x}\right] \Bigg\},
\end{align}
while for $\dot{e}_2^\alpha$, using $V=z^{1-\alpha}(1-z)^{1-\alpha}\left((1-z)^2 \frac{\kt^2}{Q^2}+ z^2 \xi\right)^\frac{\alpha}{2}$, we obtain
\begin{align}\label{eq: Sudakov for dot ECF}
\mathcal{R}^{(\text{f.o.})}_{\dot{e}_2^{\alpha}}(v,\xi)=& \mathcal{R}^{(\text{f.o.})}_{\mathring{\lambda}^\alpha}(v,\xi)+ \frac{\as\cf}{\pi} \int^1_0 \de z \Bigg\{\frac{1+(1-z)^2}{2 z} \left[ \log \frac{z^{-\frac{2}{\alpha}}}{x} \Theta \left(1-x z^\frac{2}{\alpha}\right)-\log\frac{a}{x} \Theta(a-x)
\right] \nonumber \\ &-x\frac{1-z}{z}\left[\left(\frac{1}{x}-z^\frac{2}{\alpha} \right)  \Theta \left(1-x z^\frac{2}{\alpha}\right) -\left(\frac{1}{x}-\frac{1}{a} \right) \Theta(a-x)\right]
\Bigg\},
\end{align}
where 
\begin{equation}
	a = \left(z(1-z)\right)^{-\frac{2}{\alpha}}-\frac{x z (2-z)}{(1-z)^2}.
\end{equation}
The results for the ECFs are explicitly expressed as a fixed-order calculation of the corresponding angularity, supplemented by a contribution arising from the recoil effects.
 While we are not able to find a closed form for the remaining integrals, we can study the behaviour of the eqs.~(\ref{eq: Sudakov for ECF}) and~(\ref{eq: Sudakov for dot ECF}) for small and large values of $x$. Firstly, we note that as $x\to 0$ both integrals simplify to:
\begin{align}
  \int^1_0 \de z\, \frac{1+(1-z)^2}{z}\frac{1}{\alpha} \log{(1-z)}=\frac{1}{\alpha}\left(\frac{5}{4}-\frac{\pi^2}{3}\right).
\end{align}
Thus, as expected, we recover the same massless limit. We also note that eqs.~(\ref{eq: Sudakov for ECF}) and~(\ref{eq: Sudakov for dot ECF}) explicitly show that ECFs and angularities for massless quarks start to differ beyond NLL. 
In contrast, as $x\to +\infty$, all the integrals in eq.~(\ref{eq: Sudakov for ECF}) and in~(\ref{eq: Sudakov for dot ECF}) vanish, so that the ECFs in the dead-cone region have the same behaviour, including the constant, as their angularity counterparts.

\subsection{The case of \softdrop~jets}\label{sec:FO-SD}

We now perform the fixed-order calculation for groomed jets. We focus on $\softdrop$  with $\beta=0$, i.e.\ mMDT and consider the cumulative distribution at  $\order \as$. At this order, the jet consists of the massive quark and a gluon and the radiator reads~\footnote{Henceforth, quantities calculated for groomed jets, such as cumulative distributions and radiators, will have a horizontal bar label on top.}
\begin{align}
\bar{\mathcal{R}}^{\text{(f.o.)}}_V&=\frac{\as}{2\pi}\int^1_0 \de z\int^{Q^2}_0 \frac{\de \kt^2}{\kt^2+z^2 m^2}  P_{gq} (z,\kt^2)\Theta(V(\kt,\eta)-v)\,\Theta\left(\text{min}(z,1-z)-\zc\right)\nonumber\\
&= \frac{\as}{2\pi}\int^{1-\zc}_{\zc} \de z\int^{Q^2}_0 \frac{\de \kt^2}{\kt^2+z^2 m^2}  P_{gq} (z,\kt^2)\,\Theta(V(\kt,\eta)-v).
\end{align}
We explicitly perform the analytical calculation for $\lambda^\alpha$ and then briefly comment about the other cases.  
We focus on the groomed region, $v<\zcut$, and after trading the $\kt$ integral with an integral over $\theta=\frac{\Delta R}{R_0}$, we have
\begin{align}\label{eq:lambda-groomed-start}
\bar{\mathcal{R}}^{\text{(f.o.)}}_{{\lambda}^\alpha}&=\frac{\as}{2\pi} \left[\int^{1}_{\left(\frac{v}{\zcut} \right)^\frac{2}{\alpha}} \frac{\de \theta^2}{\theta^2+\xi} \int^{1-\zc}_{\zc} \de z P_{gq}(z,\bar{k}_t^2) +\int^{\left(\frac{v}{\zcut} \right)^\frac{2}{\alpha}}_{\left(\frac{v}{1-\zcut} \right)^\frac{2}{\alpha}} \frac{\de \theta^2}{\theta^2+\xi} \int^{1-\zc}_{\frac{v}{\theta^\alpha}} \de z P_{gq}(z,\bar{k}_t^2)\right],
\end{align}
where $\bar{k}_t= z \theta Q$. The above integrals can be calculated analytically, and the result is expressed in terms of hypergeometric functions and dilogarithms, which are functions of $v$, $\xi$, and $\zcut$. However, because of the presence of an additional scale, the grooming scale set by $\zc$, we encounter more regions than in the ungroomed case.
To better illustrate this, we resort again to the Lund plane representation and consider a heavy-flavour jet with grooming, shown in fig.~\ref{fig:lund-plane}, right. The region in light pink is groomed away. 
As for ungroomed jets, the case $\alpha=1$ is depicted.
First, we note that we have a transition region between the ungroomed and groomed regimes that occurs when $\lambda^\alpha \simeq \zcut$. If we assume $\zcut > \xi^\frac{\alpha}{2}$, this transition happens in a region where the initiating parton can be considered massless. 
Secondly, as in the ungroomed case, we have the region $v>\xi^\frac{\alpha}{2}$ where we expect the radiator to be the same as in an mMDT jet initiated by a massless quark. However, the simultaneous presence of grooming and the dead cone introduces an additional transition at $v= \zc \xi^\frac{\alpha}{2}$. Below this value, the (logarithmically enhanced) phase space closes, and we expect the radiator to exhibit no logarithmic dependence on $v$ whatsoever. 
Since we want to analyse the dead cone, we are primarily interested in the region  $\zcut \xi^\frac{\alpha}{2}<\lambda^\alpha< \xi^\frac{\alpha}{2}$.
To simplify our expression, it is interesting to examine the
small-$\zcut$ limit of the eq.~(\ref{eq:lambda-groomed-start}) in this
region.
We find
\begin{subequations}
\begin{align}\label{eq:lambda-groomed-start-small-zcut}
 \bar{\mathcal{R}}^{\text{(f.o.)}}_{{\lambda}^\alpha} &\simeq \mathcal{R}^{\text{(f.o.)}}_{{\lambda}^\alpha} + \frac{\as \cf}{\pi}\Bigg[- \frac{1}{\alpha}\log^2 \frac{v}{\zcut} +\frac{x\zcut^{\frac{2}{\alpha}}}{1+x\zcut^{\frac{2}{\alpha}}} + \frac{\alpha}{2}\log\left(1+x\zcut^{\frac{2}{\alpha}}\right)-\frac{\alpha}{2}{\rm Li}_2\left(-x\zcut^{\frac{2}{\alpha}}\right)\Bigg]\\
  &\simeq \mathcal{R}^{\text{(f.o.)}}_{{\lambda}^\alpha} - \frac{\as
    \cf}{\alpha\pi}\log^2 \frac{v}{\zcut} + \mathcal{O}(\zcut).
\end{align}
\end{subequations}
On the second line, we have used the fact that in the region
of interest, $\zcut \xi^\frac{\alpha}{2}<\lambda^\alpha<
\xi^\frac{\alpha}{2}$, $\zcut\ll 1$ implies $x
\zcut^\frac{2}{\alpha}\ll 1$, with $x$ introduced in eq.~(\ref{eq:x-ratio}).
Up to power corrections in $\zcut$, the $\mathcal{O}(\as)$ constant
contribution, independent of $v, \xi,$ and $\zcut$, is therefore the
same for the ungroomed and groomed cases.
Since the constant in this calculation is purely coming from hard-collinear
radiation which is retained by mMDT up to power corrections in
$\zcut$, this result could have been anticipated on physical grounds. 
Furthermore, we have numerically checked that for $\zcut=0.1$ the finite-$\zcut$ corrections in eq.~(\ref{eq:lambda-groomed-start}) are below 5\%, both for $\alpha=1$ and $\alpha=2$.
This applies for all observables considered in this study. We will exploit this simplification in the next section when we discuss the transition-point corrections to the NLL resummation.
\section{Resummed calculations for ungroomed observables}\label{sec: resum}
In the previous section, we discussed in detail the fixed-order structure of cumulative distributions for the various observables introduced in section~\ref{sec:obs}. 
Here we aim to present the all-order calculation that allows for the resummation of large logarithms appearing due to the multiscale nature of jet substructure. 
We start with ungroomed jets and then move on to \softdrop~jets, which are considered in section~\ref{sec:groomed}.

Before presenting our calculations, let us state the accuracy we work at. As we already know from the fixed-order calculations, the expression for the radiator can include double logarithms of the observable and single logarithms of the quark mass (in the limit of small $x$) or single logarithms of $v$ and double logarithms of $\xi$ (in the limit of large $x$). 
Our calculation resums logarithms of $v$ and logarithms of $\xi$, in each of the regions defined by the hierarchy between $v$ and $\xi$, at NLL accuracy.
This results in a further simplification. As we  discussed before, all previously introduced ECFs and angularities have the same NLL behaviour. Therefore, at our accuracy level, we need to compute just one radiator:
\begin{equation}
\mathcal{R}_V(v,\xi) \xrightarrow{\text{NLL}} R(v,\xi).
\end{equation}
The  NLL cumulative distribution for jet shapes is closely related to the one for event shapes, see e.g.~\cite{Catani:1992ua,Banfi:2004yd}. It can be written as a sum over the partonic channels $\delta$:
\begin{equation}\label{eq:CAESAR}
  \begin{split}
    \Sigma_\mathrm{res}(v) &= \sum_\delta \Sigma_\mathrm{res}^\delta(v)\,,\,\,\text{with} \\  
    \Sigma_\mathrm{res}^\delta(v) &= \frac{1}{\sigma_0}\int \de \mathcal{B_\delta}
    \frac{\mathop{\de \sigma_0^\delta}}{\mathop{\de \mathcal{B_\delta}}} \exp\left[-\sum_{l\in\delta}
      R_l^\mathcal{B_\delta}(v)\right]\mathcal{S}^\mathcal{B_\delta}(v)\mathcal{F}^\mathcal{B_\delta}(v)\mathcal{H}^{\delta}(\mathcal{B_\delta})\,, 
  \end{split}
\end{equation}
where $\frac{\mathop{d\sigma_0^\delta}}{\mathop{d\mathcal{B_\delta}}}$ is the fully differential Born cross section for the partonic channel $\delta$ and $\mathcal{H}$ implements the kinematic cuts applied to the Born phase space $\mathcal{B}$. 
In eq.~(\ref{eq:CAESAR}), the normalisation $\sigma_0$ denotes the total cross section at the Born level. The radiators $R_l$ describe emissions collinear to each hard leg $l$, $\mathcal{F}$ denotes the multiple emission function, which for the additive observables (such as ECFs and jet angularities), is simply given by
$\mathcal{F}= e^{-\gamma_E R^\prime}/\Gamma(1+R^\prime)$,
with $R^\prime=\partial R/\partial L$, with $L=-\log v$, and
$R=\sum_{l\in \delta} R_l$.
The soft function
$\mathcal{S}$ implements the non-trivial aspects of colour evolution, as well as non-global logarithms~\cite{Dasgupta:2001sh}. In this paper, we work in the limiting case of a small jet radius and systematically neglect power corrections in $R_0$. This greatly simplifies the colour structure of the resummation because the contribution from soft and large-angle gluons drops out. We discuss non-global logarithms in section~\ref{sec: NGL}.
This resummation formalism has recently been implemented in the \textsc{CAESAR}  resummation plugin to \textsc{Sherpa} event generator~\cite{Gerwick:2014gya,Reichelt:2021eru} and found many phenomenological applications~\cite{Marzani:2019evv,Baberuxki:2019ifp,Baron:2020xoi,Caletti:2021oor,Reichelt:2021svh,Knobbe:2023ehi,Gehrmann-DeRidder:2024avt,Chien:2024uax}.
The focus of this work is not to provide (yet) resummed predictions for the jet phenomenology, but to study the  effects due to the presence of non-zero quark masses. Therefore, we focus on the radiator for a massive leg $l=b$ and compute it in the quasi-collinear limit.

\subsection{Radiator for massive quarks}\label{sec:radiator_quasi-coll}
The radiator in the quasi-collinear limit is closely related to the fixed-order calculations presented in section~\ref{sec:FO}. Exploiting the results of ref.~\cite{Ghira:2023bxr}, we can write it as
\begin{equation} \label{eq: R(v)}
R_b(v,\xi)=\int^{1}_0 \de z \int^{Q^2}_{z^2 m^2} \frac{\de \kt^2}{\kt^2} P_{gb}\left(z,\kt^2-z^2 m^2\right)\frac{\as^{\text{CMW}}(\kt^2)}{2\pi} \Theta\left(\left(\frac{\kt^2}{Q^2}\right)^\frac{\alpha}{2} z^{-(\alpha-1)}-v\right),
\end{equation}
where, as before, $Q=p_t R_0$. Note that we have shifted the integration variable $\kt$ and dropped all mass dependence in the $\Theta$ function. As we have already pointed out, although this is fully legitimate at NLL, kinematic mass-effects in the observable definition can be sizeable.
For this reason, we partially account for them through a matching prescription that is described in section~\ref{sec: transition 5-4}. 
The running coupling is taken in the Catani-Marchesini-Webber (CMW) scheme \cite{Catani:1990rr}:
\begin{equation}
\as^{\text{CMW}}(\kt^2)= \as(\kt^2)\left(1+\as(\kt^2)\frac{K^{(n_f)}}{2\pi}\right),
\end{equation}
with:
\begin{equation}
K^{(n_f)}=C_\text{A}\left(\frac{67}{18}-\frac{\pi^2}{6}\right)-\frac{5}{9}n_f.
\end{equation}
In order to fully take mass effects  into account, the coupling is considered in the decoupling scheme:
\begin{equation}\label{eq:alphas_dec}
\as(\kt^2)= \as^{(5)}(\kt^2) \Theta(\kt^2-m^2)+\as^{(4)}(\kt^2) \Theta(m^2-\kt^2).
\end{equation}
On the Lund plane in fig.~\ref{fig:lund-plane}, the flavour threshold is indicated by a red dotted  horizontal line.
We  write our results for the $b$-quark radiator $R_b(v)$,  as a function of the five-flavour coupling evaluated at the hard scale $\as\equiv\as^{(5)}(Q^2)$. We therefore express both $\as^{(5)}(\kt^2)$ and $\as^{(4)}(\kt^2)$ in terms of $\as$:
\begin{subequations}
\begin{align}
\as^{(5)}(\kt^2)=& \frac{\as}{1-\nu^{(5)}}\left(1-\as \frac{\beta_1^{(5)}}{\beta_0^{(5)}}\frac{\log{(1-\nu^{(5)})}}{1-\nu^{(5)}}\right),\\
\as^{(4)}(\kt^2)=& \frac{\as}{1-\nu^{(4)}-\delta_{54}}\left(1-\as \frac{\beta_1^{(4)}}{\beta_0^{(4)}}\frac{\log{\left(1-\nu^{(4)}-\delta_{54}\right)}}{1-\nu^{(4)}-\delta_{54}}\right),
\end{align}
\end{subequations}
where
\begin{align}\label{eq:nu-def}
\nu^{(n_f)}&= \as \beta_0^{(n_f)}
              \log{\left(\frac{Q^2}{\kt^2}\right)}, \qquad\text{with } n_f=4,5,\\
\delta_{54}&= \as \beta_0^{(5)}\log{\left(\frac{Q^2}{m^2}\right)}-\as\beta_0^{(4)}\log{\left(\frac{Q^2}{m^2}\right)} = \left(\beta_0^{(5)}-\beta_0^{(4)}\right)\ell_\xi.
\end{align}
In the above, we introduced the one- and two-loop coefficients of the QCD $\beta$-function:
\begin{equation}
\beta_0^{(n_f)}= \frac{11 \ca-2 n_f}{12\pi}, \qquad \beta_1^{(n_f)}=\frac{17 \ca^2-5 \ca n_f-3\cf n_f}{24\pi^2}.
\end{equation}
All integrals appearing in eq.~(\ref{eq: R(v)}) can be evaluated analytically. 
Details of the calculation, as well as explicit NLL results for the generic angular exponent $\alpha>0$, are given in appendices~\ref{app:resum} and~\ref{app. Res UG}. However, it is useful to inspect here its fixed-coupling expression
\begin{subequations}\label{eq: Sudakov fixed coupling}
\begin{align}
  R_b^\text{(f.c.)}(v,\xi)
   \overset{v>\xi^{\frac{\alpha}{2}}}&{=}
    \frac{\as \cf}{\pi}\left(\frac{1}{\alpha}\log^2{v}+\frac{3}{2\alpha}\log{v}\right); \\
   \overset{v<\xi^{\frac{\alpha}{2}}}&{=}
    \frac{\as \cf}{\pi}\Bigg(\log \xi \log v-\frac{\alpha}{4}\log^2{\xi} +\left(\frac{3}{4}-\frac{\alpha}{2}\right)\log{\xi} + \log v\Bigg).
\end{align}
\end{subequations}
At fixed coupling, the radiator receives two different contributions. In the first line, the result is the same as in the massless case, while for $v< \xi^\frac{\alpha}{2}$ the resummed exponent feels the presence of the dead cone. 
Note that this is in full agreement with our fixed-order analysis, see e.g.\ the eqs.~(\ref{eq:lambda-x}) and~(\ref{eq:lambda-circle-x}).
As discussed in the appendix~\ref{app. Res UG}, the corrections due to 
$\as$ running introduce further transitions at the flavour threshold.  
Also, let us note that, although eq.~(\ref{eq: Sudakov fixed coupling}) is continuous in $v= \xi^\frac{\alpha}{2}$, its derivative is not. 
This in turn gives rise to differential distributions that are discontinuous at the transition. In section~\ref{sec: transition 5-4} we discuss how these effects can be smoothed, but before doing that let us address non-global logarithms.

\subsection{Non-global logarithms} \label{sec: NGL} 
As it was shown in ref.~\cite{Dasgupta:2001sh}, jet shapes are affected by non-global logarithms at NLL accuracy. The first partonic configuration that gives rise to this effect involves two soft gluons where a gluon with momentum $k_1$  outside the jet emits a softer gluon with momentum $k_2$ inside the jet. Although this is  a large-angle configuration, such contribution does not vanish in the small-$R_0$ limit~\cite{Banfi:2010pa}. The resummation of non-global logarithms is highly non-trivial (for recent progress see~\cite{Banfi:2021owj,Banfi:2021xzn,Becher:2023vrh}). However, because they originate from large-angle emissions, we can expect their structure to be affected by the presence of the quark mass only by power corrections, which we neglect.

The aim of this section is to verify these considerations at $\mathcal{O}(\as^2)$ accuracy level. To do so, we consider the squared matrix element associated with the emission of two soft gluons of a massive quark anti-quark dipole. In the strongly ordered limit ($k_1\gg k_2$), we have~\cite{Czakon:2011ve,Czakon:2014oma}:
\begin{equation} \label{eq: matrix-double gluon}
|\mathcal{M}|^2\simeq \left[4\cf^2 w_{ab,1} w_{ab,2}+ 2\cf \ca w_{ab,1}\left( w_{a1,2}+w_{b1,2}-w_{ab,2}\right)\right] |\mathcal{M}_0|^2,
\end{equation}
with 
\begin{equation}
w_{ij,l}= \frac{p_i\cdot p_j}{(p_i\cdot k_l)(p_j\cdot k_l)}-\frac{p_i^2}{2(p_i\cdot k_l)^2}-\frac{p_j^2}{2(p_j\cdot k_l)^2},
\end{equation}
and $p_a, p_b$ the momenta of the massive hard partons. Here $\mathcal{M}$ denotes the original amplitude and $\mathcal{M}_0$ the truncated amplitude without soft gluons. Because the anti-$\kt$ algorithm behaves as a rigid cone in the soft limit,  there are no clustering effects at NLL~\cite{Banfi:2005gj} and we can focus on the non-Abelian $\cf \ca$ contribution to eq.~(\ref{eq: matrix-double gluon}). 
In the soft and quasi-collinear limit, after adding real and virtual contributions together, we find 
\begin{align} \label{eq: S second order}
S(v,\xi)=~&1-\left(\frac{\as }{\pi}\right)^2 \cf \ca\int^1_0 \frac{\de z_1}{z_1} \int^1_0 \frac{\de z_2}{z_2} \int^{2\pi}_0 \frac{\de \phi}{2\pi} \int^{\infty}_1 \frac{\de \theta_1^2}{\theta_1^2+\xi} \int^1_0 \frac{\de \theta^2_2}{\theta_2^2+\xi} \nonumber \\
&\times\frac{\theta_1 \theta_2 \cos{\phi} +\xi }{\theta_1^2+\theta_2^2-2\theta_1\theta_2 \cos{\phi}}\left(1-\frac{\xi}{\theta_1^2+\xi}\right)\Theta\left(z_1-z_2\right)\Theta\left(z_2 \theta_2^\alpha-v\right),
\end{align}
where $z_1, z_2$ are energy fractions and $\theta_1, \theta_2$ are the angles with respect to the jet direction (rescaled with the jet radius $R_0$) and $\phi$ is the azimuth, see also~\cite{Caletti:2023spr}. 
The upper limit on $\theta_1$ only introduces a power correction, therefore we can set it to infinity.
The mass-dependent term in the second line of eq.~(\ref{eq: S second order}) yields power corrections in both the observable and the mass, hence it can be disregarded. Moreover, setting $\theta_2=1$ in the step function $\Theta(z_2 \theta_2^\alpha-v)$ allows us to extract the leading contribution. All integrations can be performed and the result reads
\begin{equation} \label{eq: S(v)}
 S(v,\xi)= 1-\left(\frac{\as }{\pi}\right)^2 \cf \ca \left(\frac{\pi^2}{12} \log^2 v+ \text{NNLL}\right),
\end{equation}
which is the same as in the massless case. 
Indeed, this coefficient arises from the configuration in which both gluons are close to the jet boundary, while mass effects start to become significant when gluons are emitted at angles of the order of $\frac{m}{p_t}$. 
This calculation thus confirms our intuition that (at NLL accuracy) the resummation of non-global logarithms in the presence of massive particles can be performed in the same manner as in the massless case.
 
\subsection{Transition corrections}\label{sec: transition 5-4}
We have already pointed out that the NLL calculation for the quasi-collinear radiator is continuous through the transition point around the dead cone, but its derivative is not. 
However, we can exploit the fixed-order calculations performed in section~\ref{sec:FO} to smooth out this behaviour and thus obtain a better theoretical description of the region in the vicinity of the dead cone. These correction factors reproduce the $\order\as$ constant contribution, as well as power corrections in $x$, which we exponentiate in order to approximate contributions beyond NLL. Starting from eq.~(\ref{eq:CAESAR}), we define
\begin{equation}\label{eq:sigma_b_w_corr}
\Sigma_V^b(v,\xi)= \mathcal{S}(v) \frac{e^{-\gamma_{\text{E}} R_b'(v,\xi)}}{\Gamma\left(1+ R_b'(v,\xi)\right)} e^{-R_b(v,\xi)} \exp \left[- \mathcal{R}^{(\text{f.o.})}_V(v,\xi)+ R_b^\text{(f.c.)}(v,\xi)\right],
\end{equation}
where $\mathcal{R}^{(\text{f.o.})}_V(v,\xi)$ has been computed for various observables $V$ in section~\ref{sec:FO}, while the contribution $R_b^\text{(f.c.)}$, see eq.~(\ref{eq: Sudakov fixed coupling}), subtracts the logarithmic contributions. In this transition correction, the argument of the (two-loop) running coupling is set to ${\rm max} \left[m^2, Q^2 v^\frac{2}{\alpha}\right]$. The matched cumulative distribution that appears in eq.~(\ref{eq:sigma_b_w_corr}) is accurate at NLL and at $\mathcal{O}(\as)$, albeit in the quasi-collinear limit. 
Note that within our quasi-collinear approximation, the resummed contribution depends on the Born kinematics only through the jet transverse momentum, so we have dropped this explicit dependence.
\subsection{Resummation for groomed observables}\label{sec:groomed}
In this section, we briefly outline the modification to our resummed formula in the case of \softdrop~jets. Groomed jet shapes have been studied in detail, both theoretically and experimentally. \softdrop~jets are under remarkable theoretical control because the grooming procedure reduces the impact of soft emissions at large angles, simplifying the colour structure of the resummation and eliminating non-global logarithms. We work in the limit where $\zcut$ is small, so that we can neglect the power-correction, but $\log \zcut$ is not too large, so that we do not need to systematically resum these contributions. 
We present analytical results at NLL for general values of the exponent $\beta \ge 0$, even though for numerical studies we focus on the mMDT ($\beta=0$) case. 
\softdrop~jets initiated by massive quarks have been studied in~\cite{Lee:2019lge,Caletti:2023spr}. The quasi-collinear radiator with grooming is a straightforward generalisation of eq.~(\ref{eq: R(v)}) and reads
\begin{equation} \label{eq: R(v) SD}
\bar{R}_b(v,\xi,\zc)=\int^{1}_0 \de z \int^{Q^2}_{z^2 m^2} \frac{\de \kt^2}{\kt^2} P_{gb}\left(z,\kt^2-z^2 m^2\right)\frac{\as^{\text{CMW}}(\kt^2)}{2\pi} \Theta\left(\left(\frac{\kt^2}{Q^2}\right)^\frac{\alpha}{2} z^{-(\alpha-1)}-v\right) \Theta\left(z-\zcut \theta ^\beta \right).
\end{equation}
The integrations we need to perform are all of the same type as those for the ungroomed case, but, as mentioned earlier, we now have more regions due to the presence of the \softdrop~and dead-cone lines, as show in fig.~\ref{fig:lund-plane}, right. Details and explicit results are reported in appendix~\ref{app: SD calculations}. As for the ungroomed case, we provide here the radiator at fixed-coupling; for $v<\zc$ we have:
\begin{subequations}\label{eq: Sudakov fixed coupling-SD}
\begin{align}
\bar{R}_b^\text{(f.c.)}(v,\xi)\overset{v>\xi ^\frac{\alpha}{2}}&{=}  \frac{\as \cf}{\pi}\left(
\frac{\beta}{\alpha(\alpha+\beta)}\log^2{v}+\frac{2 \log \zc \log v-\log^2 \zc}{\alpha+\beta}+\frac{3}{2\alpha}\log v \right); \\
\overset{\zc \xi ^\frac{\alpha+\beta}{2}<v<\xi ^\frac{\alpha}{2}}&{=}\frac{\as \cf}{\pi} \left(-\frac{1}{\alpha+\beta}\log^2 \frac{v}{\zc} + \log \xi \log v-\frac{\alpha}{4}\log^2 \xi +\left(\frac{3}{4}-\frac{\alpha}{2}\right)\log \xi+\log v\right); \\
\overset{v<\zc \xi ^\frac{\alpha+\beta}{2}}&{=} \frac{\as \cf}{\pi} \left(\frac{\beta}{4} \log^2 \xi +\log \zc \log \xi+\frac{3}{4}\log \xi +\log \zc +\frac{\beta}{2} \log \xi \right),
\end{align}
\end{subequations}
which is in agreement with the fixed-order calculations performed for the mMDT groomed jets in section~\ref{sec:FO-SD}.
We note that in the last region the radiator freezes to a constant, leading to a vanishing differential distribution. This happens even if the observable has no explicit mass dependence. It is a genuine dynamical effect due to the simultaneous presence of the dead cone and the grooming. 
We also note a peculiar feature of the \softdrop~calculation for jets initiated by massive quarks. In the second line of eq.~(\ref{eq: Sudakov fixed coupling-SD}) we find a double logarithm of the observable $v$ with a coefficient that does not vanish when $\beta \to 0$. This might come as a surprise because jet-shapes distributions measured on mMDT jets are known to exhibit a single logarithmic behaviour. However, such contribution only exists in the region $ \zc \xi^\frac{\alpha+\beta}{2}<v<\xi^\frac{\alpha}{2}$ and not for asymptotically small values of $v$.

Finally, because in section~\ref{sec:FO-SD} we have shown that at $\mathcal{O}(\as)$ the ungroomed and groomed calculations give the same constants, up to $\mathcal{O}(\zcut)$ corrections, we construct the matched cumulative distribution as follows:
\begin{equation}\label{eq:sigma_b_w_corr_SD}
\bar{\Sigma}_V^b(v,\xi)=\frac{e^{-\gamma_{\text{E}} \bar{R}_b'(v,\xi)}}{\Gamma\left(1+ \bar{R}_b'(v,\xi)\right)} e^{-\bar{R}_b(v,\xi)}  \exp \left[- \mathcal{R}^{(\text{f.o.})}_V(v,\xi)+ R_b^\text{(f.c.)}(v,\xi)\right],
\end{equation}
i.e.\ we use the correction factor derived in the ungroomed case.

\section{Comparison to Monte Carlo}\label{sec:MC}

\begin{figure}
\includegraphics[width=0.5\textwidth, page=4]{figures/v-analytics-unc.pdf}
\includegraphics[width=0.5\textwidth, page=1]{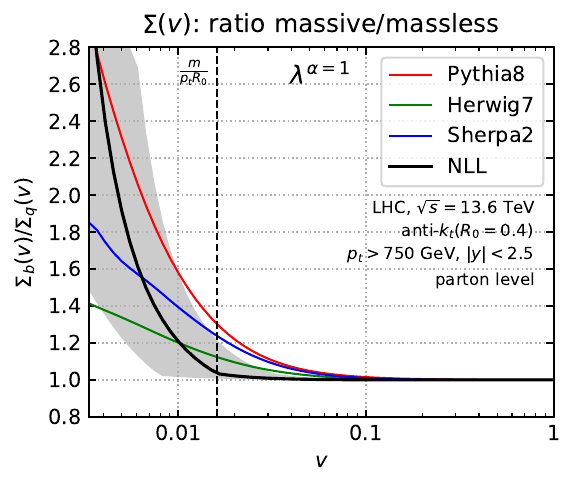}
\includegraphics[width=0.5\textwidth, page=5]{figures/v-analytics-unc.pdf}
\includegraphics[width=0.5\textwidth, page=3]{figures/v-analytics-unc.pdf}
\caption{Ratio of the cumulative distribution for $b$ jets to that for light quarks. No grooming has been applied. The left plot refers to the ECFs, the right plot to the angularities.}\label{fig:MC-comp-ungroomed}
\end{figure}

\begin{figure}
\includegraphics[width=0.5\textwidth, page=9]{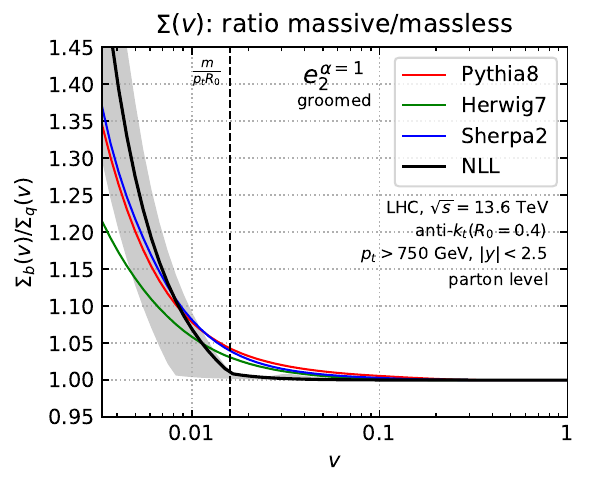}
\includegraphics[width=0.5\textwidth, page=6]{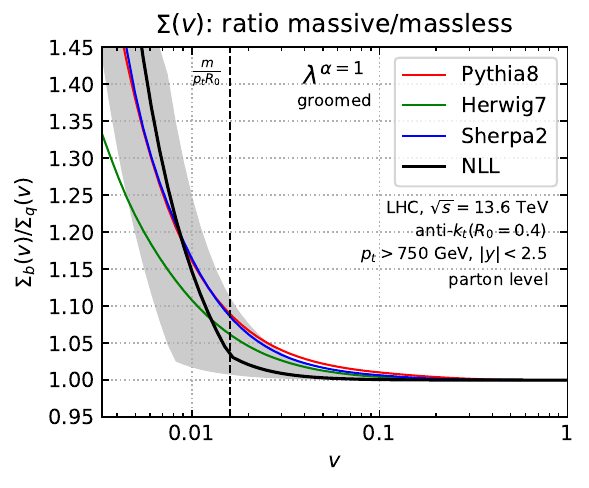}
\includegraphics[width=0.5\textwidth, page=10]{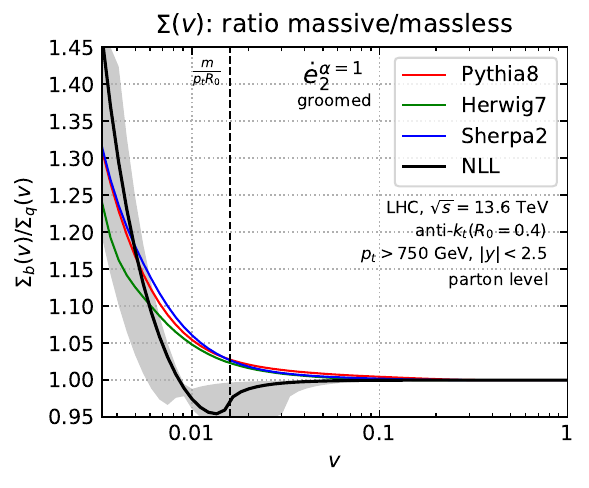}
\includegraphics[width=0.5\textwidth, page=8]{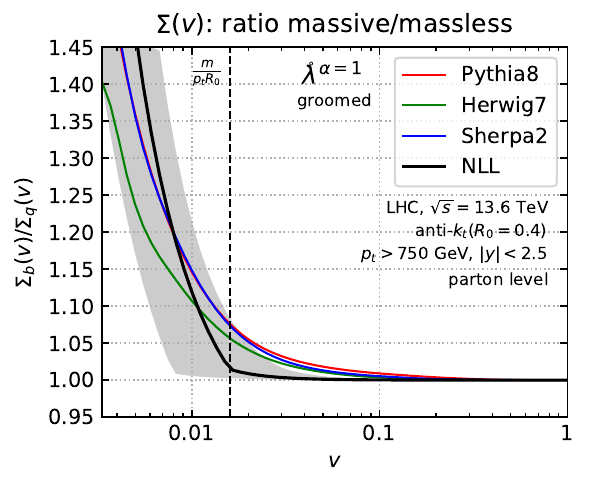}
\caption{Same observables as in figure~\ref{fig:MC-comp-ungroomed}, but for jets groomed with mMDT, with $\zc=0.1$.} \label{fig:MC-comp-groomed}
\end{figure}

We are now ready to perform some numerical studies using resummed results derived in the previous section. 
To this purpose, we compare our findings to parton-level simulations obtained with general-purpose Monte Carlo event generators, namely \pythia, \herwig and \sherpa.
Specifically, we consider the ratio of the cumulative distribution for  $b$-jets $\Sigma_V^b$ to that for  light-quark jets $\Sigma_V^q$. 
Thus, we also need to compute
\begin{equation}\label{eq:sigma_q_w_corr}
\Sigma_V^q(v,\xi)= \mathcal{S}(v) \frac{e^{-\gamma_{\text{E}} R_q'(v,\xi)}}{\Gamma\left(1+ R_q'(v,\xi)\right)} e^{-R_q(v,\xi)} \exp \left[- \mathcal{R}^{(\text{f.o.})}_V(v)+ R_q^\text{(f.c.)}(v)\right],
\end{equation}
where, to NLL accuracy, the radiator for all observables is
\begin{align}\label{eq: massless R(v)}
R_q(v,\xi)=\int^{1}_0 \de z\int^{Q^2}_{0} \frac{\de \kt^2}{\kt^2} P_{gq}\left(z\right)\frac{\as^{\text{CMW}}(\kt^2)}{2\pi} \Theta\left(\left(\frac{\kt^2}{Q^2}\right)^\frac{\alpha}{2} z^{-(\alpha-1)}-v\right),
\end{align}
and  we have introduced the standard (massless) splitting function:
\begin{equation}
P_{gq}(z)= \cf \frac{1+(1-z)^2}{z}.
\end{equation}
Note that the massless result $\Sigma_V^q(v,\xi)$ retains a dependence on $\xi$ because the integral over the running coupling is performed in the decoupling scheme, eq.~(\ref{eq:alphas_dec}). 
Furthermore, the argument of the running coupling in the transition correction is set to $Q^2 v^\frac{2}{\alpha}$.
Finally, it should be noted that the non-global contribution $\mathcal{S}$ cancels out in the ratio. The groomed version of eq.~(\ref{eq:sigma_q_w_corr}), $\bar \Sigma_V^q(v,\xi)$, can be easily calculated using the groomed radiator for a massless quark.

We show the numerical comparison between our resummed results and Monte Carlo simulations in figure~\ref{fig:MC-comp-ungroomed}, for the ungroomed jets with $\alpha=1$, and in figure~\ref{fig:MC-comp-groomed}, for the jets groomed with mMDT. We consider high-$p_t$ (above 750~GeV) jets at central rapidities ($|y|<2.5$). Jets are defined with the anti-$k_t$ algorithm, with $R_0=0.4$. 
On the left we show the ECFs, $e_2^\alpha$ at the top and $\dot{e}_2^\alpha$ at the bottom; on the right we show two of the angularities, $\lambda^\alpha$ at the top and $\mathring{\lambda}^\alpha$ at the bottom. We exclude the dotted versions of the angularities from this comparison because the presence of the end-point makes our calculation inadequate. We also refrain from showing $\mathring{\lambda}^\alpha_0$ distribution because at $\mathcal{O}(\as)$ it coincides with $\lambda^\alpha$. For the NLL calculation, the transverse momentum is $p_t=750$~GeV, with $\as(p_t R_0)=0.1$.

All resummed results are supplemented with an uncertainty band. This is obtained by replacing the argument of the strong coupling $p_t R_0 \to x_{R} \, p_t R_0$ and by rescaling the observable $v \to x_v v$ while  adding appropriate counterterms to mantain NLL accuracy. In order to provide a rough assessment of the theoretical uncertainty on the ratio in the region around the dead-cone transition, both $x_R$ and $x_v$ are varied independently $\frac{1}{2}\leq x_{R,v}\leq 2$ and the respective uncertainties are combined in quadrature. 

All plots demonstrate that, while by construction our analytic calculation become sensitive to mass effects around the transition point $\frac{m}{p_t R_0}$, all Monte Carlo simulation show a ratio that deviates from unity at much larger observable values. This can be partially explained by the fact that for each emission the dead cone opening angle is set by the energy of emitter, while in our resummed calculations it is always fixed by the energy of the hard parton\footnote{We thank Leticia Cunqueiro Mendez for discussions on this point.}.
We note that the ratio for $\dot{e}_2^\alpha$ shows a weird dip to the left of the dead cone, which is even more pronounced when grooming is applied. This is entirely driven by the $\mathcal{O}(\as)$ correction, which is negative for this observable. On the other hand, the same correction for the angularities moves the resummed curves closer to the Monte Carlo results, both for ungroomed and groomed jets.
The theoretical uncertainty of the NLL result is rather large. On the one hand, this is reassuring because it is of the same order as the observed discrepancy between the Monte Carlo predictions. On the other hand, the large uncertainty reduces the size of the dead-cone effect. We believe that this situation can be partially improved by a complete matching to fixed-order predictions, which is work in progress. However, this also motivates the calculation of full NNLL corrections that are only partially accounted for in our current results.

\begin{figure}
\includegraphics[width=0.5\textwidth, page=3]{figures/np-corrections.pdf}
\includegraphics[width=0.5\textwidth, page=1]{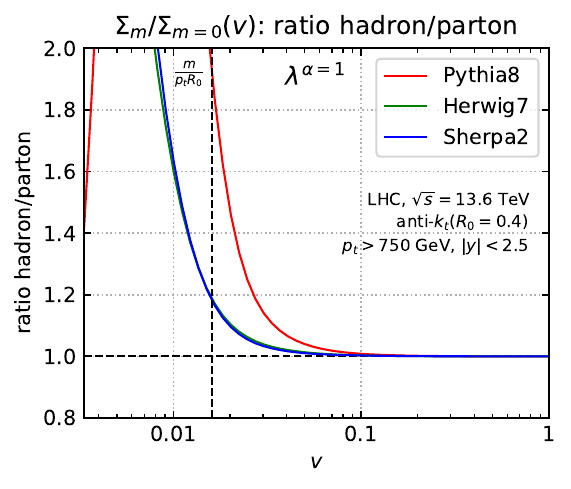}
\includegraphics[width=0.5\textwidth, page=4]{figures/np-corrections.pdf}
\includegraphics[width=0.5\textwidth, page=2]{figures/np-corrections.pdf}
\caption{Effect of non-perturbative contributions evaluated using different Monte Carlo event generators. The plots show the ratio hadron-level over parton-level for the ratio of the cumulative distributions (massive/massless) considered for the parton-level study.}\label{fig:MC-had-ungroomed}
\end{figure}

\begin{figure}
\includegraphics[width=0.5\textwidth, page=7]{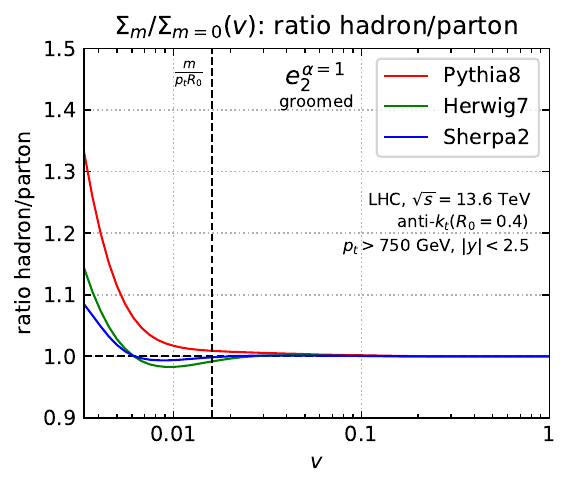}
\includegraphics[width=0.5\textwidth, page=5]{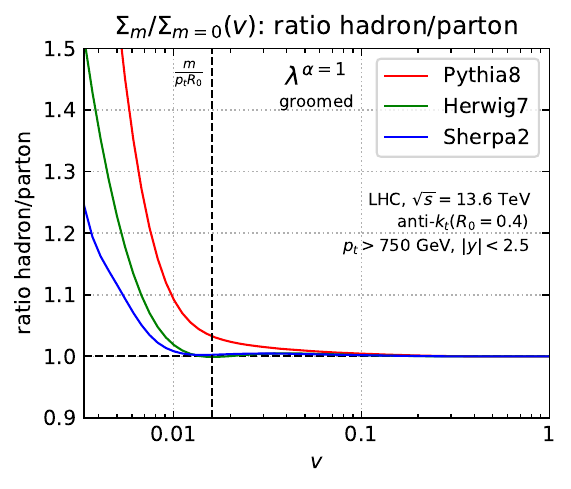}
\includegraphics[width=0.5\textwidth, page=8]{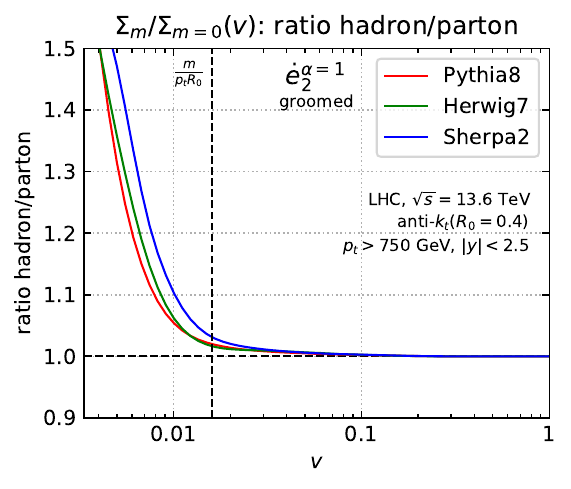}
\includegraphics[width=0.5\textwidth, page=6]{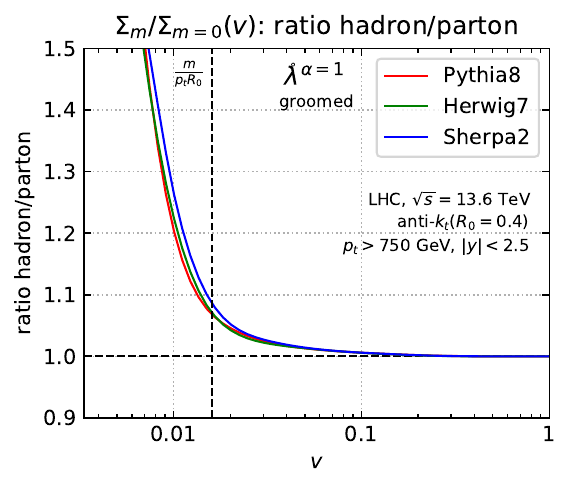}
\caption{Same observables as in figure~\ref{fig:MC-had-ungroomed}, but for jets groomed with mMDT, with $\zc=0.1$.} \label{fig:MC-had-groomed}
\end{figure}

We now turn our attention to non-perturbative correction due to the hadronisation process and the Underlying Event (UE) contribution. We perform this study by considering the ratio $\Sigma_b/\Sigma_q$ at two different stages of the event generation, namely at parton-level, i.e.\ as in the figures~\ref{fig:MC-comp-ungroomed} and~\ref{fig:MC-comp-groomed}, and at hadron-level, i.e.\ with hadronisation and the UE turned on. We keep $B$ hadrons stable and plot the ratio 
$\frac{\Sigma_b^\text{hadron}}{\Sigma^\text{hadron}_q}\Big/\frac{\Sigma_b}{\Sigma_q}$ in figure~\ref{fig:MC-had-ungroomed}, for the ungroomed case, and in figure~\ref{fig:MC-had-groomed}, for mMDT jets.
We first note that because we  selected jets with large transverse momentum, there is a wide range of $v$ for which non-perturbative effects are well under control. On the other hand, due to large $p_t$ values, the dead-cone region is pushed towards low values of $v$. As a consequence, in the case of ungroomed jets, the region where mass-effects start to become visible is affected by sizeable non-perturbative corrections.
Note that this is particularly severe for observables defined with scalar products, see for instance the bottom plots of figure~\ref{fig:MC-had-ungroomed}. On the other hand, the situation changes radically when we consider groomed jets. As shown in figure~\ref{fig:MC-had-groomed}, the non-perturbative corrections for mMDT jets with $\zcut=0.1$ remain small in the dead-cone region, especially for the shapes defined in terms of $p_{t}$ and $\Delta R$ (top plots).

To summarise, our analysis demonstrated that groomed angularities and ECFs are well-suited to study the dead cone in $b$-jets. In particular, we recommend using $\alpha=1$, as larger values will push the dead-cone region to smaller values of the observable, likely to a region that is beyond the perturbative control. 
A rather aggressive grooming, such as \softdrop~with $\beta=0$, i.e.\ mMDT, alleviates the non-perturbative contributions. 
Among the variants discussed here, the standard ones, i.e.\ $e_2^\alpha$ and $\lambda^\alpha$, are better suited to expose the dead cone, while other distributions receive mass-dependent contributions also from the scalar products in the observables' definitions. From the point of view of perturbative calculability, we have already seen that $\lambda^\alpha$ appears to be closer to the parton-level Monte Carlo predictions in the dead cone region and so this observable, measured on groomed jets, emerges from our study as the one with better analytic behaviour and resilience against non-perturbative contributions.
\section{Conclusions and Outlook}\label{sec:conclusions}
In this paper we performed a comprehensive studies of jet substructure observables measured on heavy flavour jets. 
We started our discussion by comparing different definitions of ECFs and jet angularities. While two of the variants we considered, namely $e_2^\alpha$ and $\lambda^\alpha$ are defined using the jet constituents' transverse momenta, rapidities and azimuths, the other ones make use of scalar products, either between particles' four-momenta (ECFs) or between each particle momentum and the jet axis. We found that all definitions are identical if massless particles (and massless jet axis) are considered, but differ in the massive case because scalar products introduce some mass dependence. 
From the point of view of fixed-order calculability, observables defined via scalar products are easier, essentially because the same structure appears in the observable definition and in the denominators of the squared matrix elements. For instance, we found that hypergeometric functions appear in the one-loop expression for $\lambda^\alpha$, while only logarithms appear in the equivalent calculation for scalar-product based observables, e.g. \ $\mathring{\lambda}^\alpha$. 
However, scalar products introduce an explicit mass dependence to the observable. We have studied the consequence of this fact using Monte Carlo simulations. A mass-dependent contribution in the observable definition often causes an end-point for the distribution, at low values of the observable. 
Therefore, events cannot populate these distributions at arbitrarily small values and accumulate to the right of the end-point, potentially giving rise to pronounced spikes. This effect is very pronounced for $\dot{\lambda}^\alpha$ and $\dot{\lambda}^\alpha_0$, and is much milder for the mMDT versions of $\dot{e}_2^\alpha$ and $\mathring{\lambda}^\alpha$.
Whether this is good or bad depends on one's purpose. If one is after observables that are sensitive to the mass of, say, the $b$ quark, then the presence of the mass in the observable definition acts as a cutoff and the resulting distribution exhibits a strong dependence on it. 
Thus, we could say that scalar-product based observables are very sensitive to \emph{kinematic} mass effects.
However, if one is focused on exposing the dead cone, i.e.\ the \emph{dynamical} suppression of QCD radiation collinear to a massive emitter, we would argue that explicit mass dependence should be avoided in the observable definitions.
In this regard $e_2^\alpha$ and $\lambda^\alpha$ (and, to a certain extent, $\mathring{\lambda}^\alpha_0$) are more suitable. Furthermore, for the \softdrop~ versions of these observable, the logarithmic phase-space close, resulting in (approximate) end-points of dynamical origin.

With the exception of $\dot{\lambda}^\alpha$ and $\dot{\lambda}^\alpha_0$, the observables considered in this work share  the same NLL structure, but differ beyond the NLL  accuracy level.
We calculated their cumulative distribution resumming both logarithms of the observables and logarithms of the ratio of the quark mass to the hard scale of the process, to NLL accuracy level.
Our resummed results hold for any positive value of the observable angular exponent $\alpha>0$ and \softdrop~parameters $\zcut, \beta \ge0$.
We also included fixed-order corrections that allow for a better theoretical description of the dead-cone region. 
We considered the ratio of these cumulative distributions to the ones of jets initiated by a massless quark and compared them to Monte Carlo event simulations. 
For our numerical studies, we considered $\alpha=1$ and $\beta=0$. 
While we found a good agreement between our results and Monte Carlo simulations, we consistently observe that Monte Carlo predictions show sensitivity to mass effects at larger values of the observable than suggest our calculations. 
We also found that the agreement between resummed calculations and Monte Carlo simulation is better for jet angularities than ECFs, especially if grooming was applied. 
Finally, we have studied non-perturbative corrections using Monte Carlo simulations. 
Unsurprisingly, we found that groomed jets are more resilient to hadronisation and UE contribution. 
We also found that among all observables, the variants that do not employ scalar products also show smaller non-perturbative corrections. 

In conclusion, our work  has clarified several subtleties regarding the definition of jet substructure observables in the context of heavy-flavour jets and has provided NLL results for widely used jet shapes, such as ECFs and angularities. 
The natural next step would be to turn our calculations into actual phenomenological predictions that can be compared to data. 
To this purpose, we aim to implement our results in the resummation plugin to \sherpa~\cite{Gerwick:2014gya, Reichelt:2021eru} in order to match them to full NLO theoretical predictions, with fiducial cuts, supplemented with non-perturbative corrections, as done for instance in~\cite{Marzani:2019evv,Baberuxki:2019ifp,Baron:2020xoi,Caletti:2021oor,Reichelt:2021svh,Knobbe:2023ehi}.

\section*{Acknowledgments}
The work of AG and SM is supported by the Italian Ministry of Research (MUR) under grant PRIN 2022SNA23K and by ICSC Spoke~2 under grant BOODINI.
The work of OF is supported in part by the US Department of Energy (DOE) Contract No.~DE-AC05-06OR23177, under which Jefferson Science Associates, LLC operates Jefferson Lab and by the Department of Energy Early Career Award grant DE-SC0023304.
OF also thanks the Institute for Nuclear Theory at the University of Washington for its kind hospitality and stimulating research environment. This research was supported in part by the INT's U.S. Department of Energy grant No. DE-FG02- 00ER41132.
The work of PKD is supported by European Commission MSCA Action
COLLINEAR-FRACTURE, Grant Agreement No. 101108573.
The work of GS has been supported by the European Research Council
(ERC) under the European Union’s Horizon 2020 research and innovation
programme (grant agreement No.\ 788223, PanScales).
AG and SM would like to thank IPhT Saclay for hospitality during the course of this work. 
We are grateful to Steffen Schumann and Simon Pl\"{a}tzer for providing information on implementation of heavy-quark radiation in \sherpa and \herwig MC.
Most of the simulation is conducted with computing facilities of the Galileo cluster at the Department of Physics and Astronomy of Georgia State University.

\FloatBarrier
\appendix

\section{Building blocks of the analytic resummation}
\label{app:resum}
We introduce the following notation:
\begin{equation}
\ell= -2\, \as \log v, \quad \lc=- 2 \, \as \log \zc 
\end{equation}
In order to calculate eq.~(\ref{eq: R(v)}) we exploit the Lund diagrams computing
 double integrals over rapidity and $\nu^{(n_f)}$, defined in eq.~(\ref{eq:nu-def}). As pointed out in \cite{Caletti:2023spr}, at NLL the integral over the rapidity can always be written as $ \beta_0^{(n_f)} \ell_0+c \nu$, with $\ell_0$ and $c$ independent of $\nu$. Therefore, we introduce
\begin{equation}
I^{(n_f)}(c,\ell_0,\ell_a,\ell_b)
= \frac{ \cf}{\left(\as \beta_0^{(n_f)}\right)^2}\int_{\beta_0^{(n_f)}\ell_a}^{\beta_0^{(n_f)}\ell_b} \de \nu
\frac{\as^{\text{CMW}}(\kt^2)}{2\pi} \left\lbrack \beta_0^{(n_f)} \ell_0 +c \nu\right\rbrack,
\end{equation}
with $\kt^2= Q^2 \exp\left(-\frac{\nu}{\as \beta_0^{(n_f)}}\right)$. Here, $\ell_b$, $\ell_a$ and $\ell_0$  denote a generic linear combination $\ell,\ell_\xi$ and $\lc$, while $c$ is a constant depending on the parameters $\alpha$ and $\beta$.
We evaluate this as follows:
\begin{align} \label{eq:int 4-5}
I^{(n_f)}(c,\ell_0,\ell_a,\ell_b)
&= \frac{\cf}{2\pi\as\left(\beta_0^{(n_f)}\right)^2}
 \left[\beta_0^{(n_f)} \ell_0\left(I_1^{(n_f)}(\ell_a,\ell_b)
- \frac{\beta_1^{(n_f)}}{\beta_0^{(n_f)}}\as I_2^{(n_f)}(\ell_a,\ell_b) +
\frac{\as K_\text{eff}^{(n_f)}}{2\pi} I_K^{(n_f)}(\ell_a,\ell_b)\right)\right.\nonumber\\
& \left.+c\left(J_1^{(n_f)}(\ell_a,\ell_b)
- \frac{\beta_1^{(n_f)}}{\beta_0^{(n_f)}}\as J_2^{(n_f)}(\ell_a,\ell_b) +
\frac{\as K_\text{eff}^{(n_f)}}{2\pi} J_K^{(n_f)}(\ell_a,\ell_b)\right)\right],
\end{align}
with:
\begin{equation}
K^{(n_f)}_{\text{eff}}=
\begin{cases}
K^{(5)} \quad \text{for}~n_f=5,\\
K^{(4)}-\left(\frac{\beta_1^{(5)}}{\beta_0^{(5)}}-\frac{\beta_1^{(4)}}{\beta_0^{(4)}}\right)\log{(1- \beta_0^{(5)}\ell_{\xi})}, \;\text{for}~n_f=4.
\end{cases}
\end{equation}
The integrals are given by:
\begin{align}
I_1^{(n_f)}(\ell_a,\ell_b) & = \log\frac{1-x^{(n_f)}-\beta_0^{(n_f)}\ell_{a}}{1-x^{(n_f)}-\beta_0^{(n_f)}\ell_{b}} \tag{\refstepcounter{equation}\theequation.a}, \\
I_2^{(n_f)}(\ell_a,\ell_b) & = \beta_0^{(n_f)} \frac{\ell_{b}-\ell_{a}}{(1-x^{(n_f)}-\beta_0^{(n_f)}\ell_{a})(1-x^{(n_f)}-\beta_0^{(n_f)}\ell_{b})}\nonumber  \\ 
&+ \frac{\log(1-x^{(n_f)}-\beta_0^{(n_f)}\ell_{b})}{1-x^{(n_f)}-\beta_0^{(n_f)}\ell_{b}} - \frac{\log(1-x^{(n_f)}-\beta_0^{(n_f)}\ell_{a})}{1-x^{(n_f)}-\beta_0^{(n_f)}\ell_{a}} \tag{\theequation.b},\\
I_K^{(n_f)}(\ell_a,\ell_b) & = \beta_0^{(n_f)} \frac{\ell_{b}-\ell_{a}}{(1-x^{(n_f)}-\beta_0^{(n_f)}\ell_{a})(1-x^{(n_f)}-\beta_0^{(n_f)}\ell_{b})}, \tag{\theequation.c}
\\
J_1^{(n_f)}(\ell_a,\ell_b) & = \beta_0^{(n_f)}(\ell_{a}-\ell_{b})+ (1-x^{(n_f)}) \log\frac{1-x^{(n_f)}-\beta_0^{(n_f)}\ell_{a}}{1-x^{(n_f)}-\beta_0^{(n_f)}\ell_{b}},\tag{\refstepcounter{equation}\theequation.a}\\
J_2^{(n_f)}(\ell_a,\ell_b) & = \nonumber \beta_0^{(n_f)}\frac{(\ell_{b}-\ell_{a})(1-x^{(n_f)})}{(1-x^{(n_f)}-\beta_0^{(n_f)}\ell_{a})(1-x^{(n_f)}-\beta_0^{(n_f)}\ell_{b})} \nonumber \\
+ & (1-x^{(n_f)})\left(\frac{\log(1-x^{(n_f)}-\beta_0^{(n_f)}\ell_{b})}{1-x^{(n_f)}-\beta_0^{(n_f)}\ell_{b}} 
 -   \frac{\log(1-x^{(n_f)}-\beta_0^{(n_f)}\ell_{a})}{1-x^{(n_f)}-\beta_0^{(n_f)}\ell_{a}} \right) \nonumber \\
  + & \frac{1}{2} \log^2(1-x^{(n_f)}-\beta_0^{(n_f)}\ell_{b}) - \frac{1}{2} \log^2(1-x^{(n_f)}-\beta_0^{(n_f)}\ell_{a}),\tag{\theequation.b}\\
J_K^{(n_f)}(\ell_a,\ell_b) & = \beta_0^{(n_f)} \frac{(\ell_{b}-\ell_{a})(1-x^{(n_f)})}{(1-x^{(n_f)}-\beta_0^{(n_f)}\ell_{a})(1-x^{(n_f)}-\beta_0^{(n_f)}\ell_{b})} - \log\frac{1-x^{(n_f)}-\beta_0^{(n_f)}\ell_{a}}{1-x^{(n_f)}-\beta_0^{(n_f)}\ell_{b}}.\tag{\theequation.c}
\end{align}
Finally, we have:
\begin{align}
x^{(5)} = 0, \quad
x^{(4)} = \delta_{54}.
\end{align}

The above expressions are enough to get the LL and NLL
running-coupling corrections. We still need a
few other quantities to reach NLL accuracy.
For these we can use the one-loop running coupling. Furthermore, they
can all be related to the $I_1^{(n_f)}$ integral from above. In the case of jets initiated by a massive quark,
the hard collinear splitting contribution is only relevant in the 5-flavour region, from some
logarithmic scale $\ell_b$ down to 0. It can be written as
\begin{align}
B^{(5)}\left(\ell_b\right)
= \frac{B_1}{2\pi \as \beta_0^{(5)}} \int^{ \beta_0^{(5)}\ell_b}_0 \de \nu^{(5)}~ \as^{(5)}(\kt^2)
= \frac{B_1}{2 \pi\beta_0^{(5)}} I_1^{(5)}(0,\ell_b),
\end{align}
with the usual coefficient $B_1=-\frac{3}{2} \cf$.
In the case of light jet, we could have a hard collinear contribution from the 4-flavour region from a logarithmic scale $\ell_b$ down to $\ell_\xi$.
\begin{equation}
	B^{(4)}\left(\ell_b\right)
	= \frac{B_1}{2\pi \as \beta_0^{(4)}} \int^{ \beta_0^{(4)}\ell_b}_{\beta_0^{(4)}\ell_\xi} \de \nu^{(4)}~ \as^{(4)}(\kt^2)
	= \frac{B_1}{2 \pi\beta_0^{(4)}} I_1^{(4)}(\ell_\xi,\ell_b),
\end{equation}
Additionally, the dead cone boundary furnishes an extra term for heavy flavoured jets, relevant in the 4-flavour region, from a certain
$\ell_b$ to $\ell_\xi$. It can be written as
\begin{align}\label{eq: H4}
H \left(\ell_b\right)
= \frac{H_1}{2\pi \as \beta_0^{(4)}} \int^{\beta_0^{(4)}\ell_b}_{\beta_0^{(4)}\ell_\xi} \de \nu^{(4)} \as^{(4)}(\kt^2) =\frac{H_1}{2 \pi  \beta_0^{(4)}}  I_1^{(4)}(\ell_{\xi},\ell_b),
\end{align}
with $H_1=-\cf$.
Finally, the multiple-emission NLL correction depends on $R'$, which again only
requires the 1-loop contributions $I_1^{(n_f)}$ and $J_1^{(n_f)}$.
\begin{align}
\frac{\partial}{\partial L} \left(\beta_0^{(n_f)} \ell_0 I^{(n_f)}+ c J^{(n_f)}\right)&= 
 \beta_0^{(n_f)}\Bigg[c  \left(\frac{\partial \ell_a}{\partial L} -\frac{\partial \ell_b}{\partial L} \right)+ \frac{\partial \ell_0}{\partial L} \log{\frac{1-x^{(n_f)}-\beta_0^{(n_f)}\ell_a}{1-x^{(n_f)}-\beta_0^{(n_f)}\ell_b}} \nonumber \\
+&\left(\beta_0^{(n_f)} \ell_0+c(1-x^{(n_f)})\right)\left(\frac{1}{1-x^{(n_f)}-\beta_0^{(n_f)}\ell_b}\frac{\partial \ell_b}{\partial L}-\frac{1}{1-x^{(n_f)}-\beta_0^{(n_f)}\ell_a}\frac{\partial \ell_a}{\partial L}\right) \Bigg].
\end{align}
With the above building blocks, we can construct the radiator and its derivative, which we need for the multiple-emission contribution. 

\section{Analytics for ungroomed observables}
\label{app. Res UG}
In this appendix, we report the explicit results for the resummed exponent without grooming for jets initiated by a heavy quark (denoted by the subscript $b$) by a light quark ($q$). In both cases, we include the effect of the $b$ threshold in the running of the coupling.

\subsection{$\alpha=1$}
We have two cases to consider. On the one hand, if $v>\sqrt{\xi}$ we obtain:
\begin{align}
R_b=&R_q= I^{(5)}\left(1,0,0,\ell\right)+B^{(5)}(\ell)\tag{\refstepcounter{equation}\theequation.a} \\
R_b'=& R_q'= \frac{\cf}{\pi} \frac{\ell}{1- \beta_0^{(5)}\ell}\tag{\theequation.b} \\
\end{align}
On the other hand, if $v<\sqrt{\xi}$:
\begin{align}
R_b=& I^{(5)}\left(1,0,0,\ell_\xi\right)+I^{(4)}\left(0,\ell_\xi,\ell_\xi,\ell\right)++B^{(5)}(\ell_\xi)+H(\ell) \tag{\refstepcounter{equation}\theequation.a},\\
R_b'=& \frac{\cf}{\pi } \frac{\ell_{\xi}}{1-\delta_{54}-\beta_0^{(4)}\ell}\tag{\theequation.b},\\
R_q=& I^{(5)}\left(1,0,0,\ell_\xi\right)+I^{(4)}\left(1,0,\ell_\xi,\ell\right)+B^{(5)}(\ell_\xi)+ B^{(4)}(\ell) \tag{\refstepcounter{equation}\theequation.a},\\
R_q'=& \frac{\cf}{\pi } \frac{\ell}{1-\delta_{54}-\beta_0^{(4)}\ell}\tag{\theequation.b}.
\end{align}

\subsection{$\alpha>1$}
In this case, we have to analyse three different regimes. First, for $ v>\sqrt{\xi}$ we find:
\begin{align}
R_b=&R_q= I^{(5)}\left(1,0,0,\frac{\ell}{\alpha}\right)+I^{(5)}\left(\frac{1}{1-\alpha}, \frac{\ell}{\alpha-1},\frac{\ell}{\alpha},\ell\right)+B^{(5)}\left(\frac{\ell}{\alpha}\right),\tag{\refstepcounter{equation}\theequation.a}\\
R_b'=& R_q'= \frac{\cf}{\pi \beta_0^{(5)}(1-\alpha)} I_1^{(5)}\left(\ell,\frac{\ell}{\alpha}\right)\tag{\theequation.b}.
\end{align}
Second, for  $\xi^{\frac{\alpha}{2}}<v<\sqrt{\xi}$:
\begin{align}
R_b=&R_q= I^{(5)}\left(\frac{1}{1-\alpha},\frac{\ell}{\alpha-1},\frac{\ell}{\alpha},\ell_\xi\right)+I^{(5)}\left(1,0,0,\frac{\ell}{\alpha}\right)+I^{(4)}\left(\frac{1}{1-\alpha},\frac{\ell}{\alpha-1},\ell_\xi,\ell\right)+B^{(5)}\left(\frac{\ell}{\alpha}\right),\tag{\refstepcounter{equation}\theequation.a}\\
R_b'=& R_q'= \frac{\cf}{\pi \beta_0^{(4)}\left(\alpha-1\right)} I_1^{(4)}\left(\ell_\xi,\ell\right)+\frac{\cf}{\pi \beta_0^{(5)}\left(\alpha-1\right)} I_1^{(5)}\left(\frac{\ell}{\alpha},\ell_\xi\right) \tag{\theequation.b}.
\end{align}
Third, for $v<\xi^{\frac{\alpha}{2}}$:
\begin{align}
R_b=& I^{(5)}\left(1,0,0,\ell_\xi\right)+I^{(4)}\left(\frac{1}{1-\alpha},\frac{\ell}{\alpha-1},\ell+(1-\alpha)\ell_\xi,\ell\right)\nonumber \\+& I^{(4)}\left(0,\ell_{\xi},\ell_{\xi},\ell+(1-\alpha)\ell_\xi\right)+ B^{(5)}\left(\ell_\xi\right)+H\left(\ell+(1-\alpha)\ell_\xi\right),\tag{\refstepcounter{equation}\theequation.a}\\
R_b'=& \frac{\cf}{\pi \beta_0^{(4)}(\alpha-1)} I_1^{(4)}\left(\ell+(1-\alpha)\ell_\xi,\ell\right),\tag{\theequation.b} \\
R_q=&I^{(5)}\left(1,0,0,\ell_\xi\right)+ I^{(4)}\left(\frac{1}{1-\alpha},\frac{\ell}{\alpha-1},\frac{\ell}{\alpha},\ell\right)+I^{(4)}\left(1,0,\ell_\xi, \frac{\ell}{\alpha}\right)\nonumber \\+&  B^{(5)}\left(\ell_\xi\right)+B^{(4)}\left(\frac{\ell}{\alpha}\right),\tag{\refstepcounter{equation}\theequation.a}\\
R_q'=& \frac{\cf}{\pi \beta_0^{(4)}(\alpha-1)} I_1^{(4)}\left(\frac{\ell}{\alpha},\ell\right).\tag{\theequation.b}
\end{align}

\subsection{$\alpha<1$}
Also in the present case we have to deal with three different regions.
First, if $v> \xi^{\frac{\alpha}{2}}$:
\begin{align}
R_b=&R_q= I^{(5)}\left(1,0,0,\ell\right)+I^{(5)}\left(\frac{\alpha}{\alpha-1},\frac{\ell}{1-\alpha},\ell,\frac{\ell}{\alpha}\right)+ B^{(5)}\left(\frac{\ell}{\alpha}\right),\tag{\refstepcounter{equation}\theequation.a}\\
R_b'=&R_q'= \frac{\cf}{\pi \beta_0^{(5)}(1-\alpha)} I_1^{(5)}\left(\ell,\frac{\ell}{\alpha}\right)\tag{\theequation.b}.
\end{align}
Second, if $\sqrt{\xi}<v<\xi^{\frac{\alpha}{2}}$:
\begin{align}
R_b=& I^{(5)}\left(\frac{\alpha}{\alpha-1},\frac{\ell}{1-\alpha},\ell,\ell_\xi\right) +I^{(5)}\left(1,0,0,\ell\right)+I^{(4)}\left(\frac{1}{\alpha-1},\frac{\ell}{1-\alpha}+\ell_\xi,\ell_\xi,\ell+(1-\alpha)\ell_\xi\right)\nonumber\\
+& B^{(5)}\left(\ell_\xi\right)+H\left(\ell+(1-\alpha)\ell_\xi\right)\tag{\refstepcounter{equation}\theequation.a},\\
R_b'=& \frac{\cf}{\pi \beta_0^{(4)}(1-\alpha)} I_1^{(4)}\left(\ell_\xi,\ell+(1-\alpha)\ell_\xi\right)+\frac{\cf}{\pi \beta_0^{(5)}(1-\alpha)}I_1^{(5)}\left(\ell,\ell_\xi\right),\tag{\theequation.b} \\
R_q=& I^{(5)}\left(\frac{\alpha}{\alpha-1},\frac{\ell}{1-\alpha},\ell,\ell_\xi\right) +I^{(5)}\left(1,0,0,\ell\right)+I^{(4)}\left(\frac{\alpha}{\alpha-1},\frac{\ell}{1-\alpha},\ell_\xi,\frac{\ell}{\alpha}\right)\nonumber\\
+& B^{(5)}\left(\ell_\xi\right)+B^{(4)}\left(\frac{\ell}{\alpha}\right)\tag{\refstepcounter{equation}\theequation.a},\\
R_q'=& \frac{\cf}{\pi \beta_0^{(4)}(1-\alpha)} I_1^{(4)}\left(\ell_\xi,\frac{\ell}{\alpha}\right)+\frac{\cf}{\pi \beta_0^{(5)}(1-\alpha)}I_1^{(5)}\left(\ell,\ell_\xi\right)\tag{\theequation.b}.
\end{align}
Finally, for $v<\sqrt{\xi}$:
\begin{align}
R_b=& I^{(5)}\left(1,0,0,\ell_\xi\right)+I^{(4)}\left(0,\ell_\xi , \ell_\xi, \ell\right)+I^{(4)}\left(\frac{1}{\alpha-1},\frac{\ell}{1-\alpha}+\ell_\xi,\ell,\ell+(1-\alpha)\ell_\xi\right)\nonumber \\
+&B^{(5)}\left(\ell_\xi\right)+H\left(\ell+(1-\alpha)\ell_\xi\right),\tag{\refstepcounter{equation}\theequation.a}\\
R_b'=& \frac{\cf}{\pi \beta_0^{(4)}(1-\alpha)} I_1^{(4)}\left(\ell,\ell+(1-\alpha)\ell_\xi\right)\tag{\theequation.b}, \\
R_q=& I^{(5)}\left(1,0,0,\ell_\xi\right)+I^{(4)}\left(1,0,\ell_\xi , \ell\right)+I^{(4)}\left(\frac{\alpha}{\alpha-1},\frac{\ell}{1-\alpha},\ell,\frac{\ell}{\alpha}\right)\nonumber \\
+&B^{(5)}\left(\ell_\xi\right)+B^{(4)}\left(\frac{\ell}{\alpha}\right),\tag{\refstepcounter{equation}\theequation.a}\\
R_q'=& \frac{\cf}{\pi \beta_0^{(4)}(1-\alpha)} I_1^{(4)}\left(\ell,\frac{\ell}{\alpha}\right)\tag{\theequation.b}.
\end{align}
\section{Analytics for groomed observables} \label{app: SD calculations}
We report the explicit results for the resummed exponent with grooming.
As pointed out in the main text, in this case we have to consider more regions. 
We note that if $v>\zcut$ we recover the ungroomed calculation presented in the previous section, thus we show our results only for $v<\zcut$. We will always assume that $\zc^2>\xi^\alpha, \forall \alpha$. With this in mind, let us introduce
\begin{equation}
\begin{split}
\quad v_a= \sqrt{\xi} \left(\frac{\sqrt{\xi}}{\zc}\right)^{\frac{\alpha-1}{1+\beta}},  \quad
v_b=\xi^{\frac{\alpha}{2}}, 
\quad v_c=\zc \xi^{\frac{\alpha+\beta}{2}}.
\end{split}
\end{equation}
Given that $\beta\ge0$, the hierarchy among $v_a, v_b, v_c$ depends only on $\alpha$. Specifically, $v_a = v_b > v_c$ for $\alpha = 1$, $v_a > v_b > v_c$ for $\alpha>1$ and $v_b > v_a > v_c$ for $\alpha<1$.
The scale $v_c$ corresponds to the point where the soft drop line intersects the dead cone. Consequently, this scale is absent for light quark jets.
Therefore, when computing $\RSD_q$ defined in eq.~(\ref{eq: massless R(v)}), one should only consider the hierarchy between the observable $v$ and $v_a, v_b$.
In the case of jets initiated by massive quark, when $v<v_c$ the radiator stabilizes to a constant value, and the cumulative distribution ceases to depend on $v$. This means that when $v<v_c$ the result is always the same irrespectively of $\alpha$. 
\subsection{$\alpha=1$}
In this scenario we need to analyse three cases: $v>v_a$, $v_a>v>v_c$ and $v>v_c$. 
\begin{align}
\bullet~ \zcut>&~v>v_a \nonumber\\
\RSD_b=&~\RSD_q=I^{(5)}\left(\frac{\beta}{1+\beta},\frac{\lc}{1+\beta},\lc,\ell\right)+I^{(5)}\left(1,0,0,\lc\right) + B^{(5)}\left(\ell\right)\tag{\refstepcounter{equation}\theequation.a} \\
\RSDp_b=&\RSDp_q=~\frac{\cf}{\pi (1+\beta)}\frac{\beta \ell+\lc}{1-\beta_0^{(5)}\ell}\tag{\theequation.b}
\end{align}
\begin{align}
\bullet~ v_a>&~v>v_c \nonumber\\
\RSD_b=&~I^{(5)}\left(\frac{\beta}{1+\beta},\frac{\lc}{1+\beta},\lc,\ell_\xi\right)+I^{(5)}\left(1,0,0,\lc\right) + I^{(4)}\left(-\frac{1}{1+\beta},\frac{\lc}{1+\beta}+\ell_\xi,\ell_\xi,\ell\right)\nonumber \\ ~+&B^{(5)}\left(\ell_\xi\right)+H\left(\ell\right)\tag{\refstepcounter{equation}\theequation.a} \\
\RSDp_b=&~\frac{\cf}{\pi (1+\beta)}\frac{\lc-\ell+(1+\beta)\ell_\xi}{1-\delta_{54}-\beta_0^{(4)}\ell}\tag{\theequation.b}, \\
\bullet~ v_a>&~v \nonumber\\
\RSD_q=&~I^{(5)}\left(\frac{\beta}{1+\beta},\frac{\lc}{1+\beta},\lc,\ell_\xi\right)+I^{(5)}\left(1,0,0,\lc\right) + I^{(4)}\left(\frac{\beta}{1+\beta},\frac{\lc}{1+\beta},\ell_\xi,\ell\right)\nonumber \\ ~+&B^{(5)}\left(\ell_\xi\right)+B^{(4)}\left(\frac{\ell}{\alpha}\right)\tag{\refstepcounter{equation}\theequation.a} \\
\RSDp_q=&~\frac{\cf}{\pi (1+\beta)}\frac{\beta \ell+\lc}{1-\delta_{54}-\beta_0^{(4)}\ell}\tag{\theequation.b},
\end{align}
\begin{align}
\bullet~ v_c>&~v \nonumber \\
\RSD_b=&~I^{(5)}\left(\frac{\beta}{1+\beta},\frac{\lc}{1+\beta},\lc,\ell_\xi\right)+I^{(5)}\left(1,0,0,\lc\right)\nonumber \\
+&I^{(4)}\left(-\frac{1}{1+\beta},\frac{\lc}{1+\beta}+\ell_{\xi},\ell_{\xi},\lc+\ell_{\xi}(1+\beta)\right)
+B^{(5)}(\ell_\xi)+H\left(\lc+(1+\beta)\ell_{\xi}\right)\tag{\refstepcounter{equation}\theequation.a},\\
\RSDp_b=&~0\tag{\theequation.b}.
\end{align}
\subsection{$\alpha>1$}
Here we show the analytical results for the groomed case with $\alpha>1$. There are four regions we need to consider for heavy quark jets, namely $v>v_a$, $v_a>v>v_b$, $v_b>v>v_c$ and $v_c>v$. In the case of light quark, there are only three possibilities.
\begin{align}
\bullet~\zcut>&v>v_a \nonumber \\
\RSD_b=~&\RSD_q=~I^{(5)}\left(-\frac{\alpha+\beta}{(\beta+1)(\alpha-1)},\frac{\lc}{1+\beta}+\frac{\ell}{\alpha-1},\frac{\ell}{\alpha},\lc+\frac{1+\beta}{\alpha+\beta}\left(\ell-\lc\right)\right)\nonumber \\
+&I^{(5)}\left(\frac{\beta}{1+\beta},\frac{\lc}{1+\beta},\lc,\frac{\ell}{\alpha}
\right)+I^{(5)}\left(1,0,0,\lc\right)+B^{(5)}\left(\frac{\ell}{\alpha}\right)\tag{\refstepcounter{equation}\theequation.a},\\
\RSDp_b=~&\RSDp_q= \frac{\cf}{\pi \beta_0^{(5)}(\alpha-1)} I_1^{(5)}\left(\frac{\ell}{\alpha}, \frac{1+\beta}{\beta+\alpha} \left(\ell-\lc\right)+\lc\right)\tag{\theequation.b}, 
\end{align}

\begin{align}
\bullet~ v_a>&~v>v_b \nonumber \\
\RSD_b=~ &\RSD_q=~\nonumber
 I^{(5)}\left(-\frac{\alpha+\beta}{(\beta+1)(\alpha-1)},\frac{\lc}{1+\beta}+\frac{\ell}{\alpha-1},\lc,\ell_\xi\right)
+I^{(5)}\left(-\frac{1}{\alpha-1},\frac{\ell}{\alpha-1},\frac{\ell}{\alpha},\lc\right)\nonumber \\
+&I^{(5)}\left(1,0,0,\frac{\ell}{\alpha}\right)+I^{(4)}\left(-\frac{\alpha+\beta}{(\beta+1)(\alpha-1)},\frac{\lc}{1+\beta}+\frac{\ell}{\alpha-1},\ell_{\xi},\lc+\frac{1+\beta}{\alpha+\beta}\left(\ell-\lc\right)\right) \nonumber \\
+&B\left(\frac{\ell}{\alpha}\right)\tag{\refstepcounter{equation}\theequation.a},\\
\RSDp_b=~ &\RSDp_q= \frac{\cf}{\pi \beta_0^{(4)}(\alpha-1)} I_1^{(4)}\left(\ell_\xi,\frac{1+\beta}{\beta+\alpha} \left(\ell-\lc\right)+\lc\right)+\frac{\cf}{\pi \beta_0^{(5)}(\alpha-1)} I_1^{(5)}\left(\frac{\ell}{\alpha},\ell_\xi\right)\tag{\theequation.b}
\end{align}
\begin{align}	
\bullet~v_b>&~v>v_c \nonumber \\
\RSD_b=&~I^{(5)}\left(\frac{\beta}{1+\beta},\frac{\lc}{1+\beta},\lc,\ell_\xi\right)+I^{(5)}\left(1,0,0,\lc\right)
\nonumber \\+& I^{(4)}\left(-\frac{\alpha+\beta}{(\beta+1)(\alpha-1)},\frac{\lc}{\beta+1}+\frac{\ell}{\alpha-1},\ell-(\alpha-1)\ell_{\xi},\lc+\frac{\beta+1}{\beta+\alpha}\left(\lc-\ell\right)\right)\nonumber \\
+&I^{(4)}\left(-\frac{1}{\beta+1},\frac{\lc}{\beta+1}+\ell_{\xi},\ell_{\xi},\ell-(\alpha-1)\ell_{\xi},\right)+B^{(5)}\left(\ell_\xi\right)+H\left(\ell+(1-\alpha)\ell_{\xi}\right) \tag{\refstepcounter{equation}\theequation.a},\\
\RSDp_b=&~ \frac{\cf}{\pi \beta_0^{(4)}(\alpha-1)}I_1^{(4)}\left(\ell+(1-\alpha)\ell_\xi,\lc+ \frac{\beta+1}{\beta+\alpha} \left(\ell-\lc\right)\right) \tag{\theequation.b}, \\
\bullet~v_b>&~v\nonumber \\
\RSD_q=&~I^{(5)}\left(1,0,0,\lc\right)
+ I^{(5)}\left(\frac{\beta}{1+\beta},\frac{\lc}{1+\beta}\lc,\ell_\xi\right)+I^{(4)}\left(\frac{\beta}{1+\beta},\frac{\lc}{1+\beta},\ell_\xi,\frac{\ell}{\alpha}\right)\nonumber \\
+&I^{(4)}\left(-\frac{\alpha+\beta}{(1+\beta)(\alpha-1)},\frac{\lc}{1+\beta}+\frac{\ell}{\alpha-1},\frac{\ell}{\alpha},\lc+\frac{\beta+1}{\beta+\alpha}\left(\lc-\ell\right)\right) \nonumber \\
+&B^{(5)}\left(\ell_\xi\right)+B^{(4)}\left(\frac{\ell}{\alpha}\right) \tag{\refstepcounter{equation}\theequation.a},\\
\RSDp_q=&~ \frac{\cf}{\pi \beta_0^{(4)}(\alpha-1)}I_1^{(4)}\left(\frac{\ell}{\alpha},\lc+ \frac{\beta+1}{\beta+\alpha} \left(\ell-\lc\right)\right) \tag{\theequation.b},
\end{align}
\begin{align}
\bullet~ v_c>&~v \nonumber \\
\RSD_b=&~I^{(5)}\left(\frac{\beta}{1+\beta},\frac{\lc}{1+\beta},\lc,\ell_\xi\right)+I^{(5)}\left(1,0,0,\lc\right)\nonumber \\
+&I^{(4)}\left(-\frac{1}{1+\beta},\frac{\lc}{1+\beta}+\ell_{\xi},\ell_{\xi},\lc+\ell_{\xi}(1+\beta)\right)
+B^{(5)}(\ell_\xi)+H\left(\lc+(1+\beta)\ell_{\xi}\right)\tag{\refstepcounter{equation}\theequation.a},\\
\RSDp_b=&~0\tag{\theequation.b}.
\end{align}
\subsection{$\alpha<1$}
When $\alpha<1$ we need to analyse four cases for heavy quark jets: $v>v_b$, $v_b>v>v_a$, $v_a>v>v_c$ and $v_c>v$. As before, there are only three possibilities
in the case of light quark jets.
\begin{align}
\bullet~\zcut>&v>v_b \nonumber \\
\RSD_b=~&\RSD_q=~I^{(5)}\left(-\frac{\alpha}{1-\alpha},\frac{\ell}{1-\alpha},\lc+\frac{1+\beta}{\alpha+\beta}\left(\ell-\lc\right),\frac{\ell}{\alpha}\right)\nonumber \\
+&I^{(5)}\left(\frac{\beta}{1+\beta},\frac{\lc}{1+\beta},\lc,\lc+\frac{1+\beta}{\alpha+\beta}(\ell-\lc)
\right)+I^{(5)}\left(1,0,0,\lc\right)+B^{(5)}\left(\frac{\ell}{\alpha}\right)\tag{\refstepcounter{equation}\theequation.a},\\
\RSDp_b=~&\RSDp_q=~ \frac{\cf}{\pi \beta_0^{(5)}(1-\alpha)} I_1^{(5)}\left( \frac{1+\beta}{\beta+\alpha} \left(\ell-\lc\right)+\lc,\frac{\ell}{\alpha}\right)\tag{\theequation.b}
\end{align}

\begin{align}
\bullet~ v_b>&~v>v_a \nonumber \\
\RSD_b=&~
 I^{(5)}\left(-\frac{\alpha}{1-\alpha},\frac{\ell}{1-\alpha},\lc+\frac{1+\beta}{\alpha+\beta}\left(\ell-\lc\right),\ell_\xi\right)
+I^{(5)}\left(1,0,0,\lc\right)\nonumber \\
+&I^{(5)}\left(\frac{\beta}{1+\beta},\frac{\lc}{1+\beta},\lc,\lc+\frac{1+\beta}{\alpha+\beta}\left(\ell-\lc\right)\right) \nonumber \\
+& I^{(4)}\left(-\frac{1}{1-\alpha},\frac{\ell}{1-\alpha}+\ell_\xi,\ell_\xi,\ell+(1-\alpha)\ell_\xi\right)
+B^{(5)}\left(\ell_\xi\right)+H\left(\ell+(1-\alpha)\ell_\xi\right)\tag{\refstepcounter{equation}\theequation.a},\\
\RSDp_b=&~ \frac{\cf}{\pi \beta_0^{(5)}(1-\alpha)} I_1^{(5)}\left(\lc+\frac{1+\beta}{\beta+\alpha}(\ell-\lc),\ell_\xi\right)+\frac{\cf}{\pi \beta_0^{(4)}(1-\alpha)} I_1^{(4)}\left(\ell_\xi,\ell+(1-\alpha)\ell_\xi\right)\tag{\theequation.b}\\
\RSD_q=&~I^{(5)}\left(-\frac{\alpha}{1-\alpha},\frac{\ell}{1-\alpha},\lc+\frac{1+\beta}{\alpha+\beta}\left(\ell-\lc\right),\ell_\xi\right)
+I^{(5)}\left(1,0,0,\lc\right) \nonumber \\+&I^{(5)}\left(\frac{\beta}{1+\beta},\frac{\lc}{1+\beta},\lc,\lc+\frac{1+\beta}{\alpha+\beta}\left(\ell-\lc\right)\right)+
I^{(4)}\left(-\frac{\alpha}{1-\alpha},\frac{\ell}{1-\alpha},\ell_\xi,\frac{\ell}{\alpha}\right)\nonumber \\+&B^{(5)}\left(\ell_\xi\right)+B^{(4)}\left(\frac{\ell}{\alpha}\right)\tag{\refstepcounter{equation}\theequation.a},\\
\RSDp_q=&~ \frac{\cf}{\pi \beta_0^{(5)}(1-\alpha)} I_1^{(5)}\left(\lc+\frac{1+\beta}{\beta+\alpha}(\ell-\lc),\ell_\xi\right)+\frac{\cf}{\pi \beta_0^{(4)}(1-\alpha)} I_1^{(4)}\left(\ell_\xi,\frac{\ell}{\alpha}\right) \tag{\theequation.b}
\end{align}
\begin{align}		
\bullet~v_a>&~v>v_c \nonumber \\
\RSD_b=&~I^{(5)}\left(\frac{\beta}{1+\beta},\frac{\lc}{1+\beta},\lc,\ell_\xi\right)+I^{(5)}\left(1,0,0,\lc\right)
\nonumber \\+& I^{(4)}\left(-\frac{1}{1-\alpha},\frac{\ell}{1-\alpha}+\ell_\xi,\lc+\frac{1+\beta}{\alpha+\beta}\left(\ell-\lc\right),\ell+(1-\alpha)\ell_\xi\right)\nonumber \\
+&I^{(4)}\left(-\frac{1}{1+\beta},\frac{\lc}{1+\beta}+\ell_\xi,\ell_\xi,\lc+\frac{1+\beta}{\alpha+\beta}\left(\ell-\lc\right)\right)+B^{(5)}\left(\ell_\xi\right)+H\left(\ell+(1-\alpha)\ell_{\xi}\right) \tag{\refstepcounter{equation}\theequation.a},\\
\RSDp_b=&~ \frac{\cf}{\pi \beta_0^{(4)}(1-\alpha)} I_1^{(4)}\left(\lc+\frac{1+\beta}{\alpha+\beta}\left(\ell-\lc\right),\ell+(1-\alpha)\ell_\xi\right) \tag{\theequation.b}, \\
\bullet~v_a>&~v \nonumber \\
\RSD_q=&~I^{(5)}\left(\frac{\beta}{1+\beta},\frac{\lc}{1+\beta},\lc,\ell_\xi\right)+I^{(5)}\left(1,0,0,\lc\right)
\nonumber \\+& I^{(4)}\left(-\frac{\alpha}{1-\alpha},\frac{\ell}{1-\alpha},\lc+\frac{1+\beta}{\alpha+\beta}\left(\ell-\lc\right),\frac{\ell}{\alpha}\right)\nonumber \\
+&I^{(4)}\left(\frac{\beta}{1+\beta},\frac{\lc}{1+\beta},\ell_\xi,\lc+\frac{1+\beta}{\alpha+\beta}\left(\ell-\lc\right)\right)+B^{(5)}\left(\ell_\xi\right)+B^{(4)}\left(\frac{\ell}{\alpha}\right) \tag{\refstepcounter{equation}\theequation.a},\\
\RSDp_q=&~ \frac{\cf}{\pi \beta_0^{(4)}(1-\alpha)} I_1^{(4)}\left(\lc+\frac{1+\beta}{\alpha+\beta}\left(\ell-\lc\right),\frac{\ell}{\alpha}\right) \tag{\theequation.b},
\end{align}
\begin{align}
\bullet~ v_c>&~v \nonumber \\
\RSD_b=&~I^{(5)}\left(\frac{\beta}{1+\beta},\frac{\lc}{1+\beta},\lc,\ell_\xi\right)+I^{(5)}\left(1,0,0,\lc\right)\nonumber \\
+&I^{(4)}\left(-\frac{1}{1+\beta},\frac{\lc}{1+\beta}+\ell_{\xi},\ell_{\xi},\lc+\ell_{\xi}(1+\beta)\right)
+B^{(5)}(\ell_\xi)+H\left(\lc+(1+\beta)\ell_{\xi}\right)\tag{\refstepcounter{equation}\theequation.a},\\
\RSDp_b=&0\tag{\theequation.b}.
\end{align}

\phantomsection
\addcontentsline{toc}{section}{References}
\bibliographystyle{jhep}
\bibliography{references}
\end{document}